\documentclass[preprint,superscriptaddress]{revtex4}
\usepackage{graphicx}

\newcommand{\bra}[1] {\left\langle #1 \right|}
\newcommand{\ket}[1] {\left| #1 \right\rangle}

\begin{document}

\title{Automated discovery and optimization of autonomous quantum error correction codes for a general open quantum system}

\author{Sahel Ashhab}
\affiliation{Advanced ICT Research Institute, National Institute of Information and Communications Technology, 4-2-1, Nukui-Kitamachi, Koganei, Tokyo 184-8795, Japan}
\affiliation{Research Institute for Science and Technology, Tokyo University of Science, 1-3 Kagurazaka, Shinjuku-ku, Tokyo 162-8601, Japan}

\date{\today}

\begin{abstract}
We develop a method to search for the optimal code space, induced decay rates and control Hamiltonian to implement autonomous quantum error correction (AQEC) for a general open quantum system. The system is defined by a free-evolution Lindbladian superoperator, which contains the free Hamiltonian and naturally occurring decoherence terms, as well as the control superoperators. The performance metric for optimization in our algorithm is the fidelity between the projector onto the code space and the same projector after Lindbladian evolution for a specified time. We use a gradient-based search to update the code words, induced decay matrix and control Hamiltonian matrix. We apply our algorithm to optimize AQEC codes for a variety of few-level systems. The four-level system with uniform decay rates offers a simple example for testing and illustrating the operation of our approach. The algorithm reliably succeeds in finding the optimal code in this case, while success becomes probabilistic for more complicated cases. For a five-level system with photon loss decay, the algorithm finds good AQEC codes, but these codes are not as good as the well-known binomial code. We use the binomial code as a starting point to search for the optimal code for a perturbed five-level system. In this case, the algorithm finds a code that is better than both the original binomial code and any other code obtained numerically when starting from a random initial guess. Our results demonstrate the promise of using computational techniques to discover and optimize AQEC codes in future real-world quantum computers.
\end{abstract}

\maketitle

\newpage

\section{Introduction}
\label{Sec:Introduction}

As quantum computers grow in size and computational power, allowing the implementation of increasingly long quantum algorithms, the need to protect quantum coherence is becoming increasingly pressing \cite{Ladd,Buluta}. As a result, one of the major lines of research in the field of quantum computing is the exploration of quantum error correction (QEC) \cite{Shor, Steane,Bennett,Knill,Nielsen,Terhal,Girvin}. The computational states 0 and 1, or more generally bit strings of zeros and ones, are encoded in rather complicated quantum states called the code words. The code words are designed in such a way that the quantum state distortions, i.e.~errors, that are most likely to occur naturally can be repaired with carefully designed protocols.

In the most commonly studied paradigm of QEC, the quantum system is allowed to evolve freely for some time, which results in a probabilistic leakage of the quantum state to parts of the Hilbert space outside the code space. A syndrome measurement then determines whether such an error has occurred. Importantly, it is necessary that neither the error process nor the syndrome measurement distinguish between the code words. In other words, the syndrome measurement reveals for example whether an error has occurred or not, but it does not reveal whether the encoded qubit is in state 0 or 1. A correction operation is subsequently performed, conditioned on the outcome of the syndrome measurement. The result of the measurement-correction sequence is that the system returns to its initial, possibly unknown, quantum state.

To avoid the overhead associated with performing the syndrome measurement and subsequent decision about the correction operation, researchers have developed autonomous QEC (AQEC) protocols, in which the dissipation processes that occur naturally in the engineered system are designed to always bring the system back to the code space \cite{Barnes,Sarovar}. In some sense, an automatic mechanism is devised, such that the measurement-correction sequence is performed by this mechanism without the need for a conscious observer to make a decision following every measurement or even to know what errors occurred during the dynamics.

Similarly to conventional QEC protocols, AQEC protocols are typically designed based on our existing understanding and intuition about the open-system dynamics of a given quantum system, such as a qubit array or a harmonic oscillator. A good example of this situation is bosonic codes that are designed based on our understanding of the linear dynamics of a harmonic oscillator.

However, not all physical systems used in quantum computing applications have simple structures that allow simple intuitive design of control and QEC protocols. For example, significant progress has been made in recent years in the development of qudits. In particular, the additional quantum states in a qudit, when compared to a qubit, are expected to provide advantages for QEC purposes \cite{Campbell}. Qudit systems therefore offer a great opportunity to explore new and/or enhanced classes of QEC and AQEC protocols.

In this work, we develop a method to perform a gradient-based numerical search to identify and optimize AQEC protocols for a general open quantum system. The algorithm performs three search procedures to simultaneously find (1) the optimal encoding of the computational states 0 and 1 into quantum states of the available quantum system, (2) the induced decay rates that autonomously correct errors and (3) control Hamiltonian matrix elements that sometimes need to be added to the free-evolution Hamiltonian to complete the AQEC protocol. The algorithm works by trying to maximize the fidelity of an unknown quantum state with the same state after evolution under the effect of all the processes at play, naturally occurring and controllably induced, for a set amount of time.

We analyze the performance of our numerical discovery and optimization approach for qudits with dimensions 4, 5 and 6. The algorithm consistently succeeds in finding the optimal codes in the simplest cases. The success in finding optimal codes becomes less certain as we go to more complicated decay models and/or higher dimensions. Nevertheless, the algorithm succeeds with some probability, which can be acceptable, since we can run the algorithm multiple times until we find a good solution, and we can use a good-but-not-optimal code to achieve AQEC. Furthermore, we demonstrate that the algorithm can be a powerful tool to optimize codes in cases when intuition allows us to make a good guess for an approximate AQEC code. This situation occurs naturally if the system parameters deviate slightly from a simple pattern in which a good AQEC code is known.

\section{Related Recent work}
\label{Sec:RecentWork}

There has been a significant amount of research activity on AQEC in recent years. Some studies investigated the fundamental mathematical conditions required to achieve AQEC \cite{Lihm,Lebreuilly,Kwon2025}, identifying analogues of the Knill-Laflamme conditions \cite{Knill} for AQEC.

A few studies proposed realizing AQEC dynamics in specific, carefully designed systems, especially in superconducting circuits. In Ref.~\cite{Kapit}, Kapit proposed implementing AQEC in a sustem of two coupled superconducting qubits. This proposal was realized experimentally by Li {\it et al.}~\cite{Li2024}. In Ref.~\cite{Kwon2022}, Kwon~{\it et al.} proposed realizing AQEC in a weakly anharmonic superconducting oscillator.

Other experiments include the realization of a binomial code by Gertler {\it et al.} \cite{Gertler}, a Gottesman-Kitaev-Preskill (GKP) code by Lachance-Quirion {\it et al.}~\cite{LachanceQuirion}, and a qudit GKP code by Brock {\it et al.}~\cite{Brock}, all of which used 3-dimensional cavities coupled to superconducting qubits. A few years earlier, Leghtas {\it et al.}~\cite{Leghtas} had demonstrated state stabilization of two quantum states, although coherence between the two states was not preserved in that experiment. DeBry {\it et al.}~\cite{DeBry} and Li {\it et al.}~\cite{Li2025} demonstrated AQEC in four-level codes in trapped ions.

Of particular interest to us are the recent theoretical studies that used numerical methods to discover new QEC and AQEC protocols \cite{Foesel,Wang,Zeng}. F\"osel {\it et al.}~\cite{Foesel} used reinforcement learning to find QEC codes. Wang {\it et al.}~\cite{Wang} used adjoint optimization to search for AQEC codes. We note that the philosophy of our approach is closely related to that of Ref.~\cite{Wang}. There are, however, clear differences between the two works. Wang {\it et al.}~\cite{Wang} focused on a harmonic oscillator coupled to one dissipative qubit for decay and one qubit for control, while we consider general open quantum systems. Zeng {\it et al.}~\cite{Zeng} used reinforcement learning to search for AQEC codes in a harmonic oscillator.

In the context of our work on the numerical discovery of AQEC codes, we should also mention the related work on the optimization of encoding, measurements and measurement-outcome-conditioned unitary operators in the standard measure-then-correct paradigm of QEC \cite{Reimpell,Yamamoto2005,Fletcher2007,Yamamoto2007,Kosut,Fletcher2008,PoulsenNautrup,Bausch,Olle,Su,Casanova}. References \cite{Reimpell,Yamamoto2005,Fletcher2007,Yamamoto2007,Kosut,Fletcher2008,Olle} addressed the question of whether various numerical optimization methods can efficiently find optimal QEC codes, including the code space and error correction operations. Poulsen Nautrup {\it et al.}~\cite{PoulsenNautrup} performed a numerical search that optimizes the connectivity graph in the surface code. Bausch {\it et al.}~\cite{Bausch} developed a machine-learning approach to achieve an optimal decoding strategy for noisy syndrome data when operating the surface code. References \cite{PoulsenNautrup,Olle,Su} used reinforcement learning as the optimization tool for QEC code discovery. Casanova {\it et al.}~\cite{Casanova} used Riemannian optimization and included a performance metric that favours simple QEC codes over more complicated ones.

It is also worth mentioning the theoretical studies that investigated approximate QEC \cite{Reimpell,Leung,Beny,Ng}. In this case, the Knill-Laflamme conditions are not satisfied, and hence perfect QEC cannot be achieved. One can nevertheless implement protocols to prolong the lifetime of quantum information.

\section{Formulation of the autonomous QEC code design problem as an optimization problem}
\label{Sec:Algorithm}

\subsection{Lindblad equation of motion}

The dynamical evolution of an open quantum system can, under the Markovian approximation, be described by the Lindblad equation:
\begin{equation}
\frac{d\rho}{dt} = - \frac{i}{\hbar} \left[ \hat{H} , \rho \right] + \sum_j \left\{ \hat{a}_j \rho \hat{a}_j^{\dagger} - \frac{1}{2} \hat{a}_j^{\dagger} \hat{a}_j \rho - \frac{1}{2} \rho \hat{a}_j^{\dagger} \hat{a}_j \right\},
\label{Eq:LindbladME}
\end{equation}
where $H$ is the Hamiltonian, which is responsible for unitary evolution, and $\hat{a}_j$ are the jump operators of the different decay channels. The Lindblad equation is often written with decay rates appearing in the decay term. However, we choose the form in Eq.~(\ref{Eq:LindbladME}), in which the decay rates are absorbed into the jump operators, because this convention will be more convenient when we introduce the control and induced-decay terms. Specifically, since we will search for optimal values of the rates and jump operators, it is natural to combine them, rather than spend optimization resources on redundant variables.

When no action is taken to protect the quantum information, a quantum state experiences decoherence as described by Eq.~(\ref{Eq:LindbladME}) with the naturally occurring decay channels. To implement an AQEC protocol and protect quantum states from decoherence, we add appropriately chosen terms to the Lindblad equation, both in the Hamiltonian and in additional terms that describe the controllably induced decay channels:
\begin{eqnarray}
\frac{d\rho}{dt} & = & - \frac{i}{\hbar} \left[ \left( \hat{H} + \sum_q \hat{O}_q \right) , \rho \right] +
\nonumber \\
& & \sum_j \left\{ \hat{a}_j \rho \hat{a}_j^{\dagger} - \frac{1}{2} \hat{a}_j^{\dagger} \hat{a}_j \rho - \frac{1}{2} \rho \hat{a}_j^{\dagger} \hat{a}_j \right\} + \sum_l \left\{ \hat{b}_l \rho \hat{b}_l^{\dagger} - \frac{1}{2} \hat{b}_l^{\dagger} \hat{b}_l \rho - \frac{1}{2} \rho \hat{b}_l^{\dagger} \hat{b}_l \right\},
\label{Eq:LindbladMEWithControl}
\end{eqnarray}
where $\hat{O}_q$ are Hamiltonian control operators, and $\hat{b}_l$ are controlled decay jump operators.

Since Eq.~(\ref{Eq:LindbladMEWithControl}) is linear in the density matrix $\rho$, the $n\times n$ matrix can be rearranged into an $n^2$-dimensional vector, i.e.~a one-dimensional array ($\tilde{\rho}$), such that Eq.~(\ref{Eq:LindbladMEWithControl}) is expressed as
\begin{equation}
\frac{d\tilde{\rho}}{dt} = \tilde{L} \tilde{\rho},
\end{equation}
where the operator, or superoperator, $\tilde{L}$ encodes all the information about the Hamiltonian and decay terms. This rearrangement of the matrix elements simplifies the task of solving the differential equation and optimizing the control parameters in it. For example, the flattened density matrix after time $t$ is given by
\begin{equation}
\tilde{\rho}(t) =\exp \left[ \tilde{L} t \right] \tilde{\rho}(0),
\end{equation}
where $\tilde{\rho}(0)$ is the initial state at time $t=0$.

\subsection{Defining target operation for optimal control calculation}

We are now ready to formulate the AQEC design problem as an optimization problem. We start by noting that we focus on encoding a single bit of quantum information, i.e.~a Hilbert space with two quantum states, even though encoding larger Hilbert spaces is also possible. The goal is to preserve any quantum superposition of the two code words as long as possible. This goal can be expressed in terms of aiming to implement a unitary operator, specifically the unit operator, which leaves any initial state unchanged. Since we are interested in protecting states that are in the code space at the initial time, we need an operator that acts as the unit operator on the code space but is insensitive to any input state that is outside the code space, similarly to what was done in Ref.~\cite{Ashhab2022} for optimizing two-qubit gates in larger Hilbert spaces. The operator that we need in this case is the projector on the code space:
\begin{equation}
\hat{P} = \ket{\tilde{0}} \bra{\tilde{0}} + \ket{\tilde{1}} \bra{\tilde{1}},
\end{equation}
where $\ket{\tilde{0}}$ and $\ket{\tilde{1}}$ are the code words. We emphasize here that the projector $\hat{P}$ is also an $n^2 \times n^2$ matrix, or superoperator, that operates on the flattened density matrix. As such, it has only four nonzero matrix elements when expressed in any basis that contains the code words as basis states. When applied to a density matrix, it keeps the matrix elements that correspond to the states $\left\{ \ket{\tilde{0}}, \ket{\tilde{1}} \right\}$ and their superpositions, while all other parts of the density matrix are eliminated, i.e.~they correspond to the eigenvalue zero of $\hat{P}$.

In order to define a performance metric for the numerical optimization algorithm, we imagine that we let the system evolve for a finite duration, and we choose a value for the evolution time $\tau$, which can in principle be any finite value, i.e.~$0<\tau<\infty$. We then define the fidelity between the implemented operation and the target operator $\hat{P}$ as
\begin{equation}
F = \frac{1}{4} \left| {\rm Tr} \left\{ \exp \left[ \tilde{L} \tau \right] \hat{P} \right\} \right|.
\label{Eq:Fidelity}
\end{equation}
This formula can be understood as follows: In a basis that contains the states $\left\{ \ket{\tilde{0}}, \ket{\tilde{1}} \right\}$, each one of the four nonzero terms inside the trace focuses on one of the four matrix elements of the density matrix in the code space and calculates how this term is degraded by the open-system dynamics described by $\tilde{L}$. 
The factor $1/4$ is used because the projector $\hat{P}$ has four matrix elements equal to one, with zeros everywhere else. The fidelity is equal to 1 if and only if the density matrix remains unaffected by the dynamical evolution, i.e.~it experiences decoherence-free evolution. It should be noted that Eq.~(\ref{Eq:Fidelity}) treats the four matrix elements as if they were independent of each other, even though they are not. We find this formulation convenient when treating the density matrix as a one-dimensional vector, without worrying about the fact that this vector must obey the standard conditions for a physical density matrix. It should also be noted that there is no unique definition of the fidelity for the purpose of searching for AQEC codes. In principle, any function that is equal to a certain value when the quantum state is preserved but decreases when we move away from perfect state protection will be an acceptable function in the calculations. We find the fidelity in Eq.~(\ref{Eq:Fidelity}) to be a computationally convenient choice, as it is obtained using standard matrix operations.

\subsection{Temporal evolution of the fidelity}

The fidelity $F$ is obviously equal to one when $\tau=0$, i.e.~when the quantum state has not started evolving. Unless the system parameters allow perfect state preservation, all choices for the AQEC code will have $F<1$ for any nonzero value of $\tau$. In our calculations, we generally set $\gamma\tau=1$, where $\gamma$ is the rate of the natural decay (i.e.~error) process.

Two points should be noted here in relation to the dynamics and the use of the fidelity as a performance metric. Both points are also related to the fact that, when evaluating the performance of QEC codes, it seems intuitive to compare the effective lifetimes with and without the application of the QEC protocol, rather than analyzing a fidelity function. Firstly, in a typical AQEC setting, the system quickly reaches a near-steady state that contains a mixture of the code words and (a small population of) the error states. This near-steady state then slowly decays to the infinite-time, true steady state. As a result, as far as the long-time dynamics is concerned, the fidelity does not start its decay from the value $F=1$, but from a slightly reduced value. It is therefore in general not straightforward to translate the fidelity value into a decay rate. We can extract a decay rate by considering the fidelity as a function of time. However, this task requires the evaluation of the fidelity at multiple time values, which would complicate the optimization algorithm. On the other hand, we can identify the optimal AQEC protocol by choosing a somewhat arbitrary value of $\tau$. Once we have obtained an AQEC code in this way, we can calculate the effective decay rate under the influence of the obtained protocol. Secondly, the decay channels can be different with and without the application of the AQEC protocol. For example, in this work we focus on the case where the free evolution involves only one decoherence channel, namely energy relaxation. Considering that the system under the influence of the AQEC protocol is a driven dissipative system, the effective dynamics will in general be qualitatively different from the free-evolution dynamics. The decoheherence channels can be different, as we will show with at least one example in Sec.~\ref{Sec:Results}. Furthermore, there can be multiple decoherence channels, each with its own rate, in the effective dynamics. As a result, it is not completely straightforward to make a comparison between the decoherence rates with and without the AQEC protocol based on individual fidelity values. When the decoherence rates between the two situations are different by orders of magnitude, the comparison between decoherence rates calculated from simple formulae seems to be logical. When the difference between the decoherence rates is small, say at the level of a factor of 2 or less, the comparison between rates can be misleading, depending on other details of the physical system under consideration. We therefore use the fidelity as a general purpose performance metric that is reasonably predictive and easy-to-calculate, while keeping in mind that any standard metric must be supplemented by a knowledge of the nature of the effective dynamics. We will also calculate ratios between the bare and effective lifetimes in some cases. As an estimate for the ratio between the effective decay rate under AQEC and the decay rate under free evolution, we will use the formula
\begin{equation}
\kappa = \frac{ (F(t = \beta_1/\gamma) - F(t=\beta_2/\gamma)) / (\beta_1-\beta_2) \Bigg|_{\rm AQEC}}{dF/d(\gamma t) \Bigg|_{t=0, \ \rm free \ evolution}},
\end{equation}
where $\beta_1$ and $\beta_2$ are chosen to give fidelity values that are slightly below the initial value, such that the numerator gives a good approximation for the slope of the fidelity at early times. For realistic AQEC parameters, $\kappa$ should be largely insensitive to the exact values of $\beta_1$ and $\beta_2$, in addition to being insensitive to the fact that the fidelity starts at a value below $F=1$ even at $t$ values that are much smaller than the natural decay times.

\subsection{Optimization algorithm}

\begin{figure}[h]
\includegraphics[width=14.0cm]{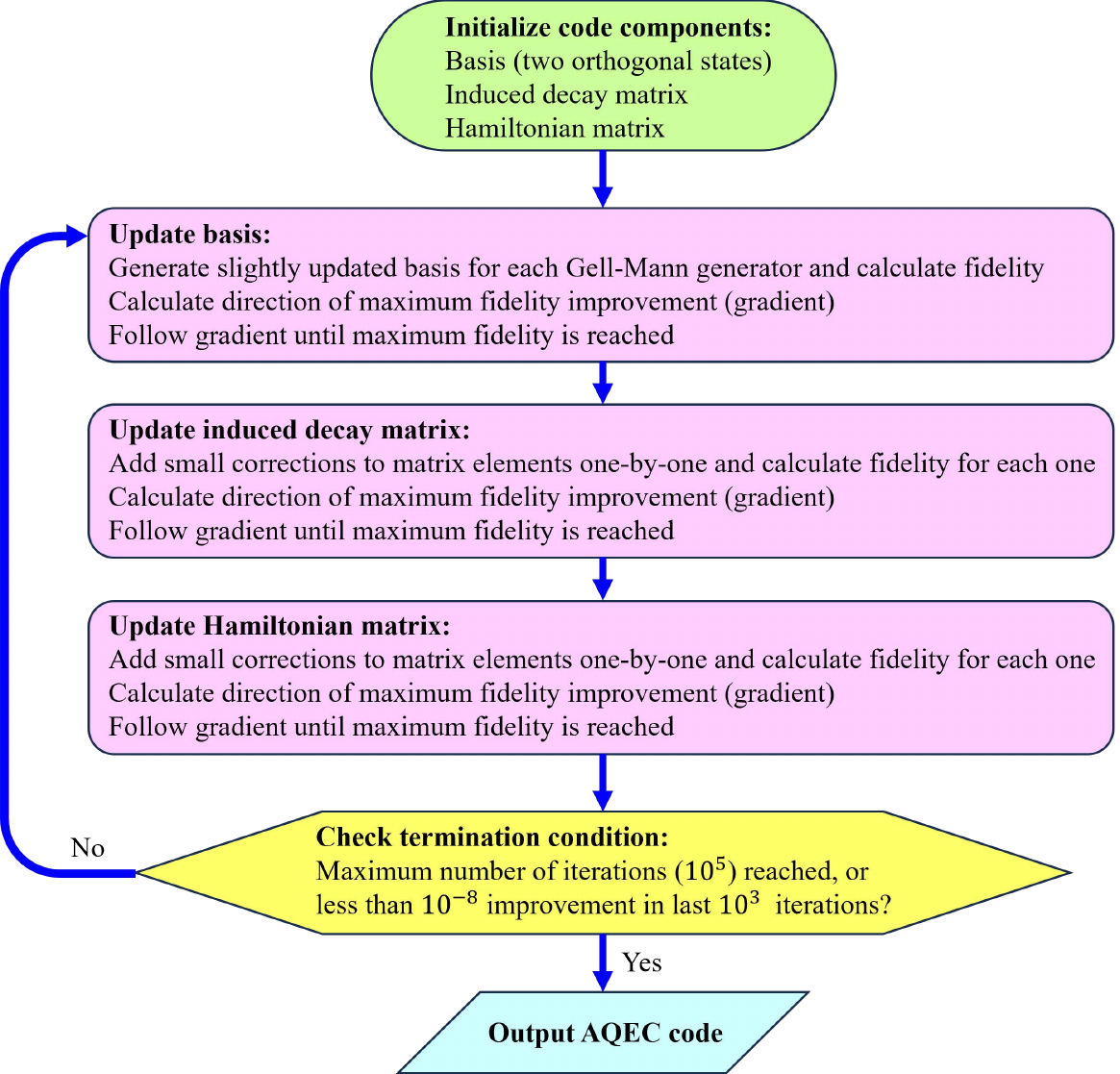}
\caption{Flowchart that summarizes the algorithm. First, the three components of the AQEC code are initialized, possibly to random initial values. These three components are then updated by following the gradients to improve the fidelity defined in Eq.~(\ref{Eq:Fidelity}). After every iteration, the two termination condition are tested. When neither condition is satisfied, the AQEC code is updated further. When one of the termination conditions is satisfied, the algorithm outputs the optimized AQEC code.}
\label{Fig:Flowchart}
\end{figure}

The optimization algorithm is illustrated in Fig.~\ref{Fig:Flowchart}. We start each instance of the search by making an initial guess for the AQEC code parameters. In most calculations, we make a random guess. As mentioned above, the AQEC code has three components: (1) the code space made up of the two code words, (2) the induced decay matrices, and (3) the controlled Hamiltonian matrix. The code space is chosen by taking two orthogonal but otherwise random vectors in the full Hilbert space. We use a single matrix $\hat{b}$ for the second component of the code. The matrices $\hat{b}$ and $\hat{O}$ are filled with complex numbers, such that each real or imaginary part of each matrix element is chosen from a uniform distribution in the range $[-0.5,0.5]$. One constraint is that $\hat{O}$ must be Hermitian. We therefore randomly generate the matrix elements above the diagonal and take their complex conjugates to calculate the matrix elements below the diagonal.

We use gradient-based techniques to optimize the three variable components of the AQEC code. We update each one of these three components by evaluating the gradient of the fidelity with respect to the relevant parameters and shifting the AQEC code parameters in the direction of the gradient. We alternate between updating the three components, such that each optimization iteration is composed of three separate update steps.

In principle, we could calculate a single, unified gradient that incorporates the parameters of all three components of the AQEC code. However, we chose to divide the step of updating the code parameters into three distinct update steps, alternating between the three and updating each one of them in each optimization iteration. There are a few reasons why we find the alternating update approach logical. For example, the code space updates are performed via applying unitary operations to the code space, while the matrices $\hat{b}$ and $\hat{O}$ are updated via simple addition. Furthermore, the optimal matrix $\hat{b}$ is expected to contain infinite matrix elements, as we will discuss in Sec.~\ref{Sec:Results}. We therefore expect some matrix elements to keep growing indefinitely but always remain infinitely far from the optimal values. These considerations make it conceptually simpler to keep the update steps separate. For comparison purposes, we repeated some of our calculations, specifically those presented in Sec.~\ref{Sec:Results}.B, using the unified-gradient approach where all three components are updated simultaneously. We found that these calculations take somewhat less time per optimization iteration but achieve a smaller fidelity improvement per iteration, such that the two approaches exhibit comparable overall convergence speeds.

An optimization iteration proceeds as follows: Taking the best available code space, we consider that any neighboring basis can be obtained from the current basis using the update unitary operator
\begin{equation}
U_{\rm update} = \exp \left\{ i \sum_m \lambda_m G_m \right\},
\label{Eq:GellMannUpdateMatrix}
\end{equation}
where $\lambda_m$ are (small) coefficients and $G_m$ are the generators of unitary operators in the full Hilbert space, i.e.~the generalized Gell-Mann matrices for the full Hilbert space. We generate a slightly modified basis for every generator, using the small infinitesimal coefficient $\lambda_m=10^{-8}$, and construct an array for the fidelity for the different possible updates. This array allows us to calculate the gradient to maximize the fidelity improvement. More specifically, we obtain a single generator that contains the optimal proportions of the different $G_m$ matrices. We can update the code words by applying to them the matrix $U_{\rm update}$ with the optimal generator multiplied by an update step size, which we initially set at $10^{-2}$. We keep moving along the direction of the gradient until the fidelity reaches a maximum and starts decreasing. When we identify that we have overshot the fidelity peak, we start zooming in on the peak region until we can identify the peak location to a precision of $10^{-8}$ in the update step size, i.e.~until an update step of $10^{-8}$ in either direction decreases the fidelity rather than increases it. We update the code words to those that correspond to the point of maximum fidelity.

For the controlled decay matrix $\hat{b}$, since it can be any $n\times n$ matrix, we take $2n^2$ possible updates for the real and imaginary parts of all matrix elements, each time adding a small increment to one of the matrix elements and calculating the fidelity with the small update. We use the increment size $10^{-6}$. Once we have the $2n^2$ values of fidelity for all the possible updates, we calculate the gradient and determine the optimal direction for updating $\hat{b}$. We then move in the optimal update direction until we reach a maximum in the fidelity. We note here that although we set a maximum for the allowed update in a single iteration ($10^4$ to any real or imaginary part of any matrix element), this condition was never needed by our numerical search algorithm. Including the condition was motivated by the fact that the optimal matrix $\hat{b}$ is expected to have some infinitely large matrix elements. When a peak in the fidelity is encountered during the update process, which is what happened in all iterations of all calculations, we start zooming in on the peak region until we can identify the maximum point location to within $10^{-8}$. A similar procedure is used to update the matrix $\hat{O}$, incorporating the constraint that the operator must be Hermitian. 

A natural question to consider here is whether the gradient-based approach is suitable for the problem of AQEC search, or whether this problem is prone to suffering from the presence of local optima that hinder the search for the optimal code. Our use of this approach was inspired by the results of Refs.~\cite{Yamamoto2005,Yamamoto2007,Kosut}, which alternated between optimizing the code space and the recovery operators in the conventional, measure-then-correct QEC paradigm. Those studies demonstrated that the search does not contain local minima, such that the gradient-based search and the alternation between updating different components of the code are good elements in a QEC code discovery algorithm. Our results in Sec.~\ref{Sec:Results} below show some indications that our search sometimes gets trapped in local optima and therefore fails to find optimal solutions. The authors of Ref.~\cite{Wang} made a similar observation and argued that the use of the average fidelity as an optimization cost function results in the appearance of local optima.

In our numerical calculations, we perform a maximum of $10^5$ optimization iterations. We terminate the algorithm early if the fidelity stagnates; specifically, if a sequence of $10^3$ iterations does not improve the fidelity by at least $10^{-8}$. The computation time exhibited some run-to-run variations, in addition to variations between different computers that we used in this work. As a rough estimate, a complete calculation with $10^5$ iterations on a four-level quantum system took about five hours on a single core of a present-day computer. For a five-level system, $10^5$ iterations took about ten hours on about 16 cores of a present-day computer running the calculation in parallel. For a six-level system, $10^5$ iterations took about 24 hours on about 25 cores. We note that the number of cores used in parallel calculations was chosen automatically by the computers, as opposed to being set manually by us.

\section{Results}
\label{Sec:Results}

In this section, we present numerical results that demonstrate the operation of the algorithm to find and optimize AQEC codes.

\subsection{Reference point: physical qubit with relaxation}

To assess the AQEC protocols analyzed later in this section, we start by establishing the reference point against which the protocols should be evaluated.

We take a physical qubit with a relaxation dynamics described by the lowering (or annihilation) operator
\begin{equation}
\hat{a} = \sqrt{\gamma} \left(
\begin{array}{cc}
0 & 1 \\
0 & 0
\end{array}
\right).
\end{equation}
An initial density matrix
\begin{equation}
\rho(0) = \left(
\begin{array}{cc}
\rho_{00} & \rho_{01} \\
\rho_{10} & \rho_{11}
\end{array}
\right)
\end{equation}
decays following the formula
\begin{equation}
\rho(t) = \left(
\begin{array}{cc}
\rho_{00} + (1-e^{-\gamma t}) \rho_{11} & e^{-\gamma t/2} \rho_{01} \\
e^{-\gamma t/2} \rho_{10} & e^{-\gamma t} \rho_{11}
\end{array}
\right).
\end{equation}
The fidelity formula in Eq.~(\ref{Eq:Fidelity}) can then be evaluated by examining how much of each matrix element survives the relaxation dynamics. Concretely, we ignore the term $(1-e^{-\gamma t}) \rho_{11}$ in the 00 matrix element, replace every $\rho_{ij}$ by 1, take the sum of all four matrix elements and divide by 4:
\begin{equation}
F = \frac{1}{4} \left( 1 + 2 e^{-\gamma t/2} + e^{-\gamma t} \right).
\label{Eq:FidelityVsDecoherenceRateTime_Relaxation}
\end{equation}
The fidelity starts off at $F=1$, decays with initial rate
\begin{equation}
\frac{dF}{dt} \Bigg|_{t=0} = - \frac{\gamma}{2}
\label{Eq:FidelityInitialRate_Relaxation}
\end{equation}
and approaches the asymptotic value $F=1/4$ in the limit $t\to\infty$.

One fidelity value that will serve as an important reference point below is the one obtained by setting $\gamma t=1$, because we use this setting in our numerical optimization calculations. With this setting, the fidelity under free evolution, i.e.~when no error-correction action is taken, is $F_0=0.64523519$.

It is also useful for later purposes to consider the case of pure dephasing. If we consider a qubit that experiences pure dephasing at rate $\gamma$ with no other decoherence channels, the density matrix decays following the formula
\begin{equation}
\rho(t) = \left(
\begin{array}{cc}
\rho_{00} & e^{-\gamma t} \rho_{01} \\
e^{-\gamma t} \rho_{10} & \rho_{11}
\end{array}
\right),
\end{equation}
which gives the fidelity
\begin{equation}
F = \frac{1}{2} \left( 1 + e^{-\gamma t} \right).
\label{Eq:FidelityVsDecoherenceRateTime_Dephasing}
\end{equation}
By comparing Eqs.~(\ref{Eq:FidelityVsDecoherenceRateTime_Relaxation}) and (\ref{Eq:FidelityVsDecoherenceRateTime_Dephasing}), we can see that the fidelity decays following different functional forms for different decoherence mechanisms.

\subsection{Ququart with uniform decay rates}

\begin{figure}[h]
\includegraphics[width=8.0cm]{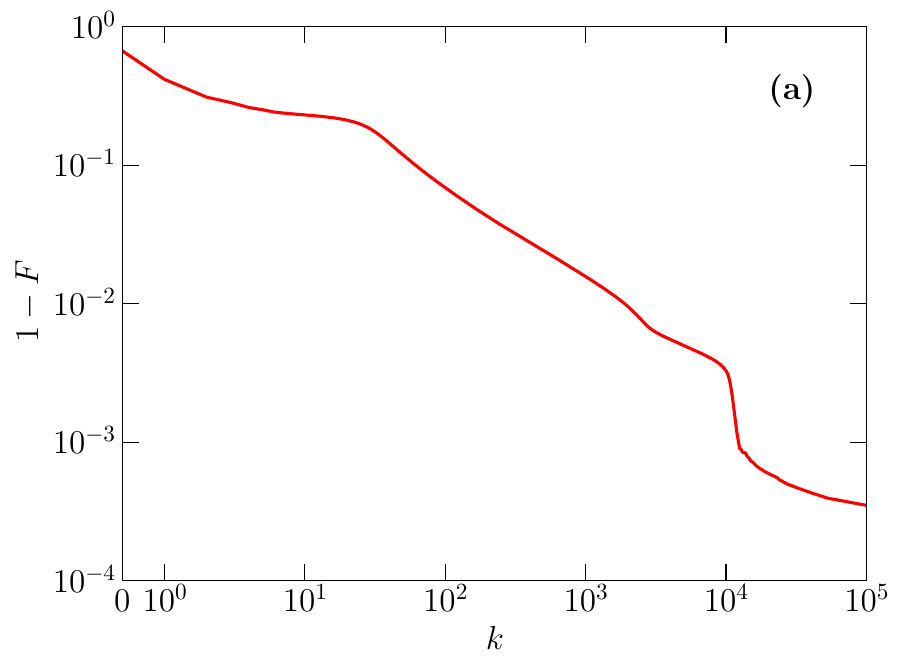}
\includegraphics[width=8.0cm]{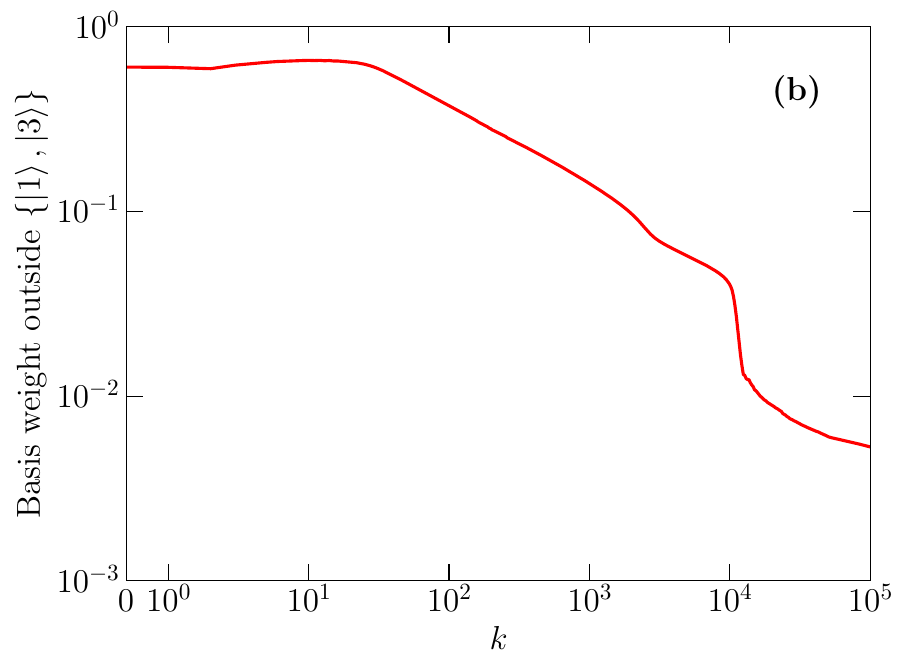}
\includegraphics[width=8.0cm]{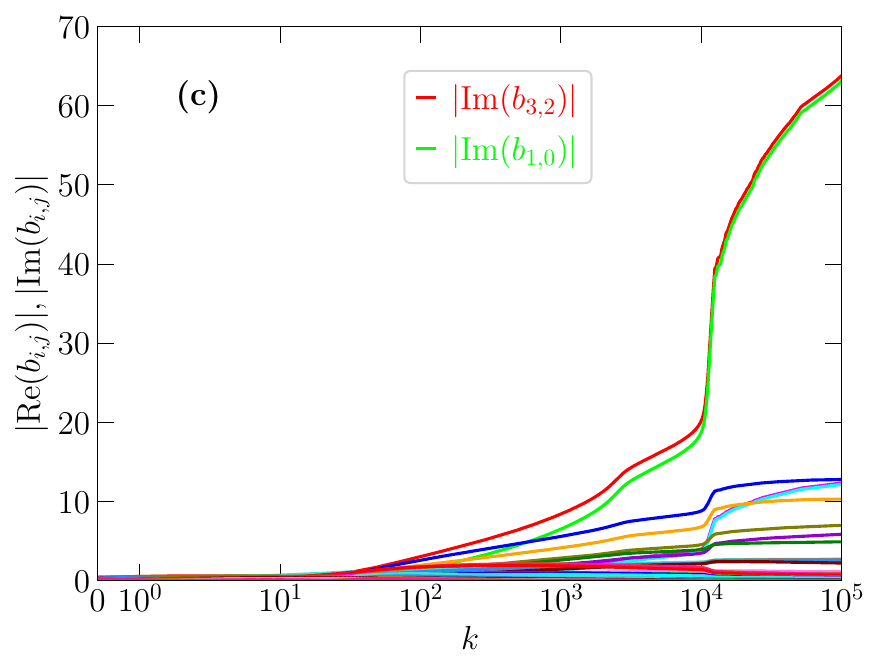}
\includegraphics[width=8.0cm]{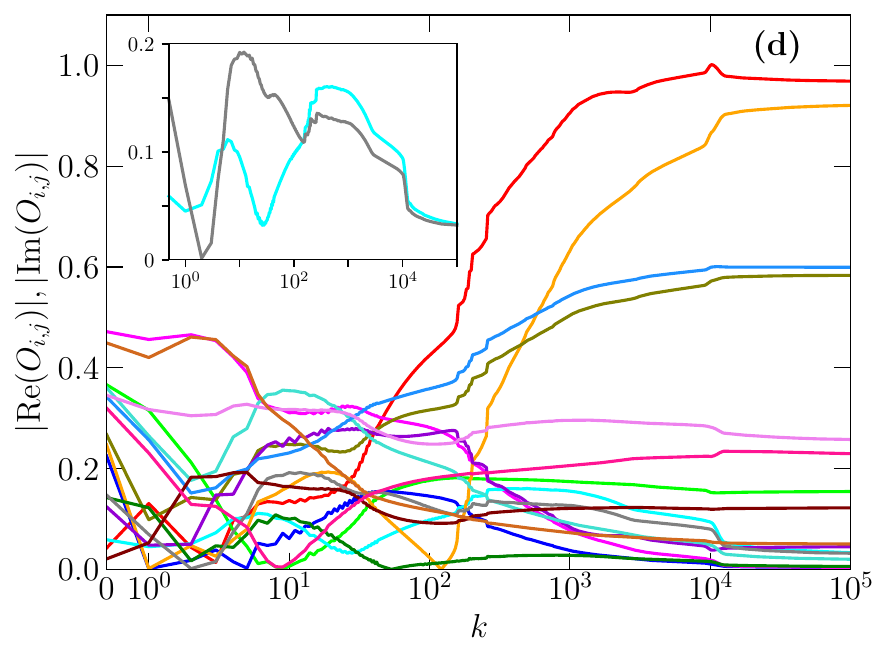}
\caption{Progression of AQEC code with optimization iteration number $k$ for 4-level system with uniform decay rates. The infidelity $(1-F)$, plotted in Panel (a), decreases steadily and approaches zero as a function of $k$, indicating that the algorithm is successful in finding a good AQEC code. The somewhat irregular behaviour of the slope suggests that the fidelity landscape has saddle-point-like features. Panel (b) shows that the code space approaches $\left\{\ket{1},\ket{3}\right\}$. In Panel (c) the legend specifies only the highest two lines. In this specific run, by coincidence, the real parts of $b_{1,0}$ and $b_{3,2}$ did not grow significantly up to $k=10^5$. In Panel (d), the inset shows the real (cyan) and imaginary (gray) parts of $O_{1,3}$. This matrix element is expected to converge to zero in the optimal code. Most other matrix elements in $\hat{O}$ are expected to have a small effect on the fidelity, as explained in Appendix \ref{Sec:AppendixOptimization}. All plots in this figure indicate that the algorithm is progressing towards the 13 code but that it has not converged yet.}
\label{Fig:ConvergenceOptimizeAll}
\end{figure}

Arguably the simplest system that allows a nontrivial form of AQEC is a four-level quantum system (ququart) with uniform decay rates down the energy level ladder $\ket{3}\rightarrow\ket{2}\rightarrow\ket{1}\rightarrow\ket{0}$. The decay process is described by the jump operator
\begin{equation}
\hat{a} = \sqrt{\gamma} \left(
\begin{array}{cccc}
0 & 1 & 0 & 0 \\
0 & 0 & 1 & 0 \\
0 & 0 & 0 & 1 \\
0 & 0 & 0 & 0
\end{array}
\right).
\end{equation}
One can intuitively guess what is almost certainly the optimal strategy, to which we will refer as the 13 code. The code space is the space spanned by the states $\ket{1}$ and $\ket{3}$. These states decay to the states $\ket{0}$ and $\ket{2}$, respectively. The induced decay matrix that brings the state after a decay event back to the original state is given by
\begin{equation}
\hat{b} = \sqrt{\Gamma} \left(
\begin{array}{cccc}
0 & 0 & 0 & 0 \\
1 & 0 & 0 & 0 \\
0 & 0 & 0 & 0 \\
0 & 0 & 1 & 0
\end{array}
\right).
\label{Eq:FourLevelInducedDecayOp}
\end{equation}
When $\Gamma\gg\gamma$, the error correction action is successful, and the effective decay rate of the fidelity is much smaller than $\gamma$. Specifically, the uncorrectable decay can be attributed to the transition $\ket{2}\to\ket{1}$ in the jump operator $\hat{a}$. If this matrix element were not present, the system would just make transitions back and forth between the code space $\left\{ \ket{1}, \ket{3} \right\}$ and the error subspace $\left\{ \ket{0}, \ket{2} \right\}$. Because the transition matrix elements in $\hat{a}$ are equal, and similarly for $\hat{b}$, the transitions between the code space and error spaces would not affect the encoded quantum state, i.e.~there would be zero decoherence of the encoded state. The uncorrectable error rate caused by the $\ket{2}\to\ket{1}$ transition can be calculated as follows: when $\Gamma\gg\gamma$, the occupation probability of the subspace $\left\{ \ket{0}, \ket{2} \right\}$ is approximately given by $\gamma/\Gamma$. If the uncorrectable decay rate out of this subspace is $\gamma$, the net uncorrectable decay rate is $\gamma^2/\Gamma$. The decoherence channel in the effective dynamics is energy relaxation from $\ket{3}$ to $\ket{1}$, i.e.~the same as in the absence of the AQEC protocol. No Hamiltonian control term $\hat{O}$ is needed in this case.

If we take the code space $\left\{ \ket{1}, \ket{3} \right\}$ and set $\Gamma/\gamma=10^6$ and $\gamma\tau=1$, we obtain the fidelity $F=0.9999985=1-1.5\times 10^{-6}$. This value is probably the best achievable value for the fidelity (at time $\tau=1/\gamma$). In fact, if we run the optimization algorithm using the above settings, the algorithm terminates immediately because of a computation error, which we believe is caused by the fact that the code that is provided as an initial guess is already at the maximum of the fidelity landscape. We note here that the fidelity can be expressed as $F=1-1.5 \gamma/\Gamma$. Two thirds of the fidelity reduction, i.e.~$\gamma/\Gamma$, can be attributed to the admixture of error states in the long-time dynamics. The rest, i.e.~$\gamma/(2\Gamma)$, results from the relaxation at rate $\gamma^2/\Gamma$ under the influence of the AQEC protocol for duration $t=1/\gamma$. The factor of 2 here is the same as the one in Eq.~(\ref{Eq:FidelityInitialRate_Relaxation}).

We now try the automated code search. As mentioned above, we start with a randomly generated basis and randomly generated matrices $\hat{b}$ and $\hat{O}$, and we set $\gamma\tau=1$. The fidelity for these random settings is typically in the range 0.25-0.5. Results from a typical run of the optimization algorithm are shown in Fig.~\ref{Fig:ConvergenceOptimizeAll}. In most runs, the fidelity reached values around $F=0.996$ after $10^4$ iterations and $F=0.99965$ after $10^5$ iterations. We did not encounter any runs that seemed to get trapped in local optima with lower fidelities. All the components of the AQEC code converged towards the optimal values described above. 

In Appendix \ref{Sec:AppendixOptimization}, we present the results of optimizing the three AQEC code components individually. By comparing the results of this subsection with those of Appendix \ref{Sec:AppendixOptimization}, we conclude that the bottleneck for the fidelity convergence is the optimization of the induced decay matrix $\hat{b}$. Furthermore, the code basis and the coherent control matrix $\hat{O}$ both keep changing as long as $\hat{b}$ is changing, indicating that these components of the AQEC code keep finding optimal values that correspond to the current value of the matrix $\hat{b}$.

It is worth making a few comments about the model treated in this section before moving on to the next model. The four-level model could be seen as a toy model that is useful as a tool to gain a basic understanding of the idea of AQEC. At the same time, it is quite conceivable that a physical system, e.g.~a superconducting circuit with a cleverly designed arrangement of Josephson junctions, can have a quantum state structure with transition rates that allow the realization of the 13 code. It is important to note here that for the 13 code to work, the $\ket{2}\to\ket{1}$ transition rate does not need to have any particular relation to the $\ket{1}\to\ket{0}$ and $\ket{3}\to\ket{2}$ transition rates. The only requirement is that the latter two be equal and that these transitions not leave a signature of the exact pair involved in the transition, e.g.~by emitting photons of different frequencies.

\subsection{Ququart with photon loss}

We now consider the lowest four energy levels of a harmonic oscillator (which can be treated as a ququart) with photon loss decay, i.e.~a decay jump operator given by the appropriately truncated harmonic oscillator annihilation operator
\begin{equation}
\hat{a} = \sqrt{\gamma} \left(
\begin{array}{cccc}
0 & 1 & 0 & 0 \\
0 & 0 & \sqrt{2} & 0 \\
0 & 0 & 0 & \sqrt{3} \\
0 & 0 & 0 & 0
\end{array}
\right).
\label{Eq:FourLevelAnnihilationOp}
\end{equation}
We assume that there are no other decoherence channels. We ran the optimal AQEC code search algorithm on this case. When we initialized the AQEC code components to random values, the calculations consistently converged to the code space $\left\{ \ket{0}, \ket{1} \right\}$, with $\hat{b}=\hat{O}=0$. In other words, the optimal protocol found by the algorithm is the one where we do nothing.

We then ask what happens if we apply the 13 code of Sec.~\ref{Sec:Results}.B in this case. Even when the decay jump operator is given by Eq.~(\ref{Eq:FourLevelAnnihilationOp}), the states $\ket{1}$ and $\ket{3}$ are stabilized effectively by the operator $\hat{b}$ in Eq.~(\ref{Eq:FourLevelInducedDecayOp}), which quickly brings back any population that decays into the error states $\ket{0}$ and $\ket{2}$, respectively. The key matrix elements to analyze are then $\rho_{02}$ and $\rho_{13}$, or equivalently $\rho_{20}$ and $\rho_{31}$. The equations governing the dynamics of these matrix elements are
\begin{eqnarray}
\dot{\rho}_{02} & = & \sqrt{3} \gamma \rho_{13} - \Gamma \rho_{02} - \gamma \rho_{02}
\nonumber \\
\dot{\rho}_{13} & = & - 2 \gamma \rho_{13} + \Gamma \rho_{02}.
\end{eqnarray}
The last term in the equation for $\rho_{02}$ arises from the relaxation of the state $\ket{2}$ into the state $\ket{1}$, which is an uncorrectable error that will make only a small contribution to the effective decoherence dynamics when $\Gamma\gg\gamma$. We therefore ignore this term. The terms with the coefficient $\Gamma$ correspond to the fast reexcitation of the error states $\ket{0}$ and $\ket{2}$ back to the code words $\ket{1}$ and $\ket{3}$, respectively. We are then left with the following situation: $\rho_{13}$ decays at the rate $2\gamma$ and is replenished at the rate $\sqrt{3}\gamma$ that appears in the first term in the equation for $\rho_{02}$. The net effect is that $\rho_{13}$ decays at the rate $(2-\sqrt{3})\gamma$. In other words, the effect of applying this AQEC code is to replace the free-evolution situation in which we have only relaxation with rate $\gamma$ by an effective situation in which we have mainly pure dephasing between the code words with a rate of roughly $\gamma/4$. Using Eq.~(\ref{Eq:FidelityVsDecoherenceRateTime_Dephasing}), the fidelity in this situation is expected to be given by $F\approx 0.88$, which is significantly higher than the fidelity $F_0\approx 0.65$. It is then surprising that the algorithm did not find the active AQEC code. This result suggests that there are local optima in the fidelity landscape.

It is worth noting here that the relatively modest improvement described in the previous paragraph, from relaxation with rate $\gamma$ to pure dephasing with rate $\gamma/4$, is unlikely to materialize in a realistic experiment, considering that the driving protocol in any realistic setup will contribute its own noise and, partially or fully, negate the improvement gained by the error correction mechanism.

It is also interesting to consider the generalization of the 13 code dynamics described above, as well as the code found in Ref.~\cite{Zeng}, to the scenario where the code space is composed of the Fock states $\ket{N}$ and $\ket{N+2}$ (or more generally $\ket{N_1}$ and $\ket{N_2}$ with $\left| N_1-N_2 \right| \geq 2$) of a harmonic oscillator. The equations of motion for the relevant off-diagonal matrix elements become
\begin{eqnarray}
\dot{\rho}_{N-1,N+1} & = & \sqrt{N(N+2)} \gamma \rho_{N,N+2} - \Gamma \rho_{N-1,N+1} - N \gamma \rho_{N-1,N+1}
\nonumber \\
\dot{\rho}_{N,N+2} & = & - (N+1) \gamma \rho_{N,N+2} + \Gamma \rho_{N-1,N+1}.
\end{eqnarray}
Following the same argument as above, we find that the effective dynamics is that of pure dephasing in which $\rho_{N,N+2}$ decays at the rate $\left(N+1-\sqrt{(N+1)^2-1}\right)\gamma$, which can be approximated as $\gamma/\left(2\sqrt{N}\right)$ for large $N$. One can therefore, in theory, reduce the effective decay rate to arbitrarily small values by using very large photon numbers, keeping in mind that the induced decay rate $\Gamma$ must now satisfy the condition $\Gamma\gg N\gamma$. The reason behind the suppression of the effective decay rate is that, in the limit $N\rightarrow\infty$, the decay rates of the states $\ket{N}$ and $\ket{N+2}$ become essentially equal, which leads to the simple scenario described in Sec.~\ref{Sec:Results}.B.

\subsection{Photon loss in higher-dimensional qudit}

\begin{figure}[h]
\includegraphics[width=8.0cm]{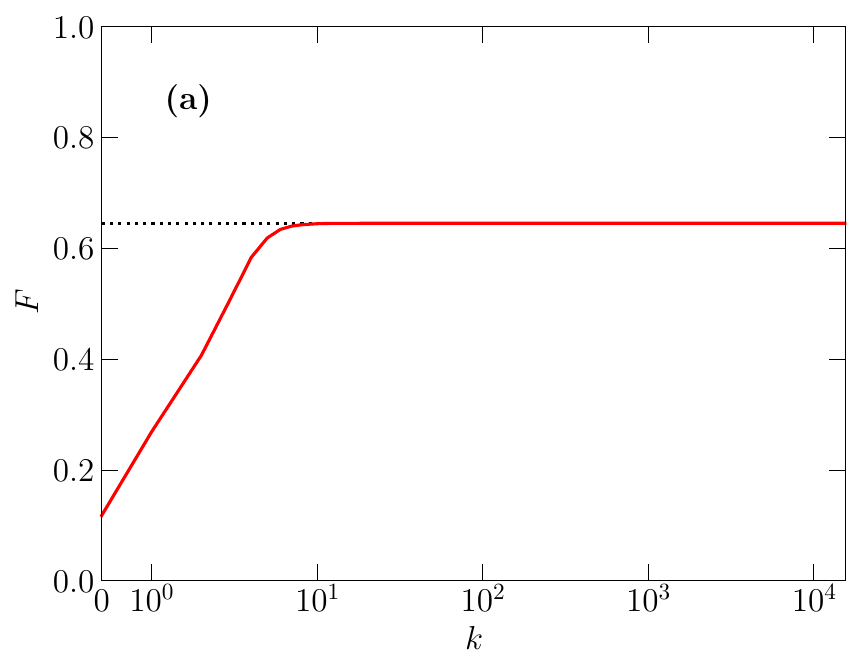}
\includegraphics[width=8.0cm]{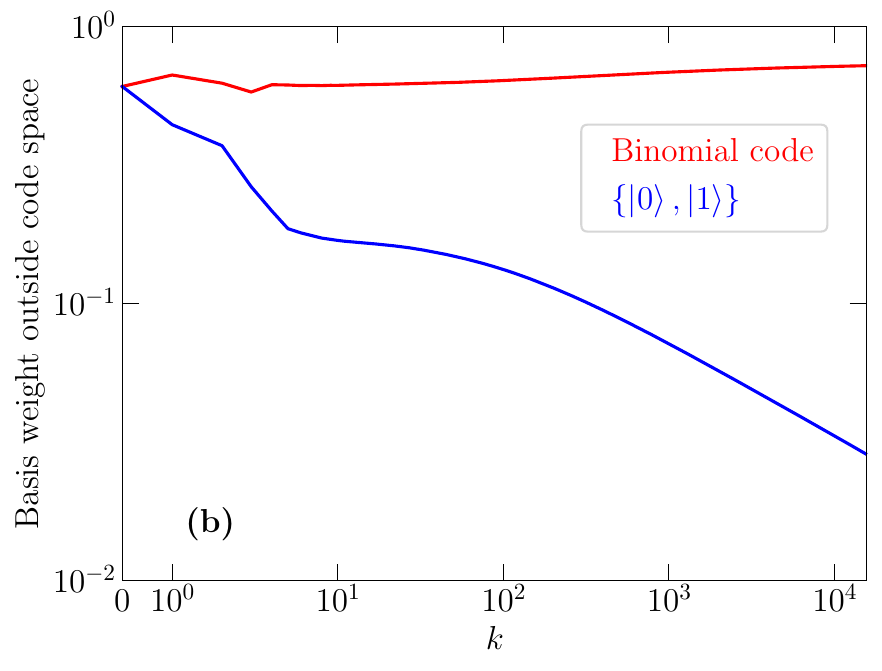}
\includegraphics[width=8.0cm]{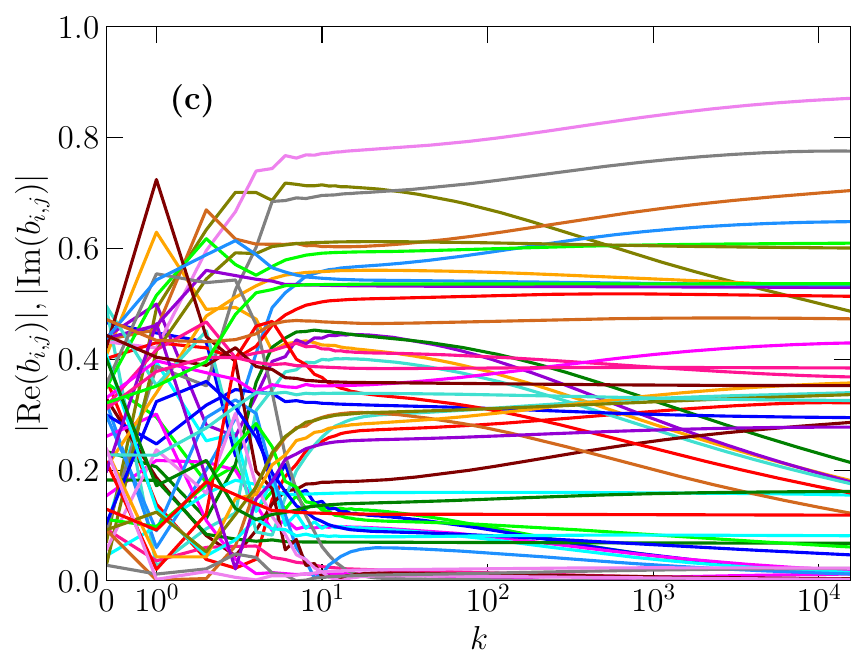}
\includegraphics[width=8.0cm]{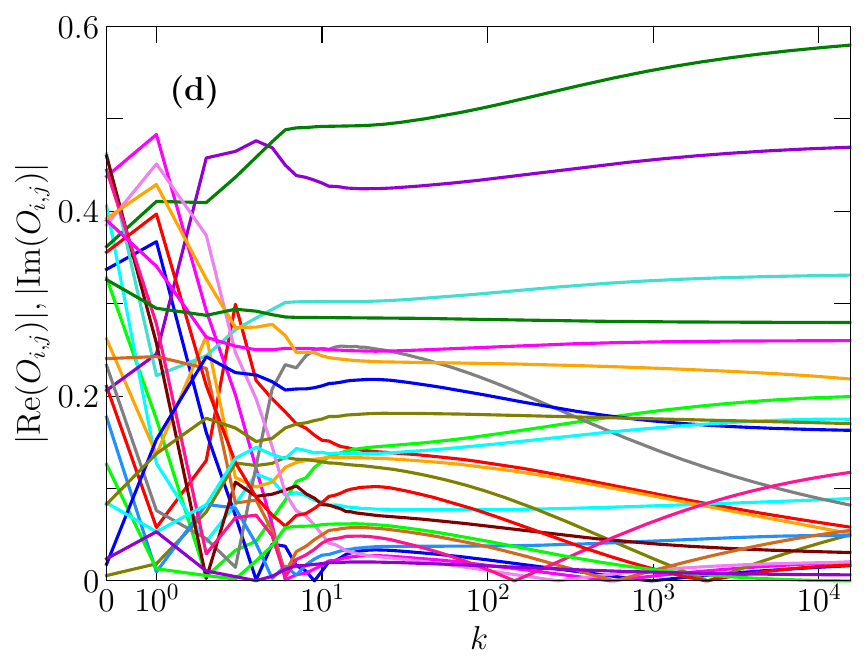}
\caption{Progression of AQEC code with optimization iteration number $k$ for 5-level system with photon loss decay. This figure shows an instance in which the algorithm failed to produce a nontrivial AQEC code. The fidelity converges to the free-evolution value $F_0$ (shown as the horizontal dotted line in Panel a). Panel (b) shows the probability for the code to be outside two code spaces that can be intuitively guessed as candidate code spaces, namely the space of the binomial code and the space $\left\{ \ket{0}, \ket{1} \right\}$. The code space gradually approaches $\left\{ \ket{0}, \ket{1} \right\}$. All the plots in this figure indicate that the algorithm is converging towards the trivial code of effectively doing nothing. This figure is representative of most runs of the algorithm. Figure \ref{Fig:Convergence5} shows one of the relatively rare instances where the algorithm found a nontrivial code.}
\label{Fig:Convergence501}
\end{figure}

\begin{figure}[h]
\includegraphics[width=8.0cm]{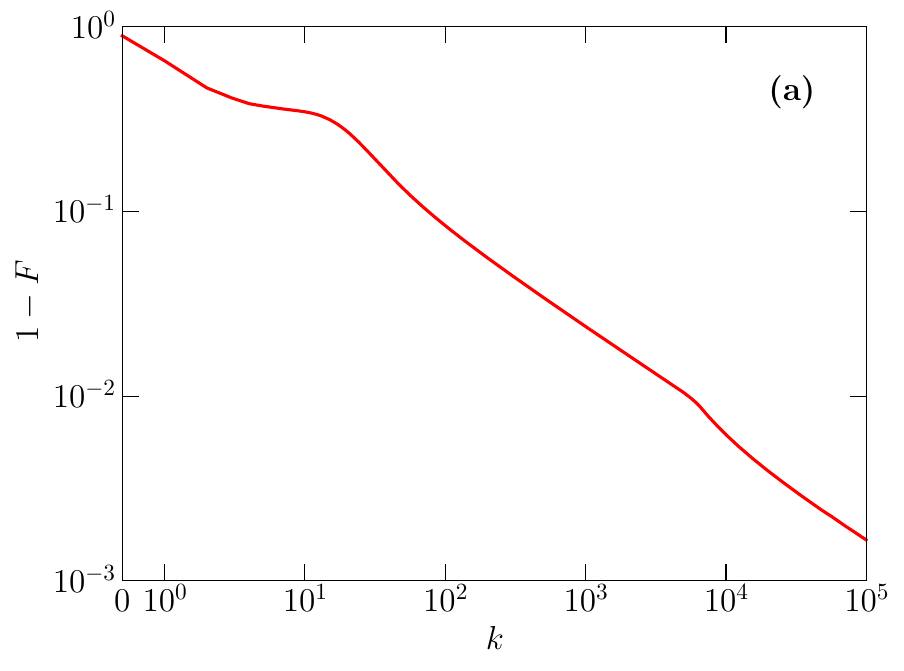}
\includegraphics[width=8.0cm]{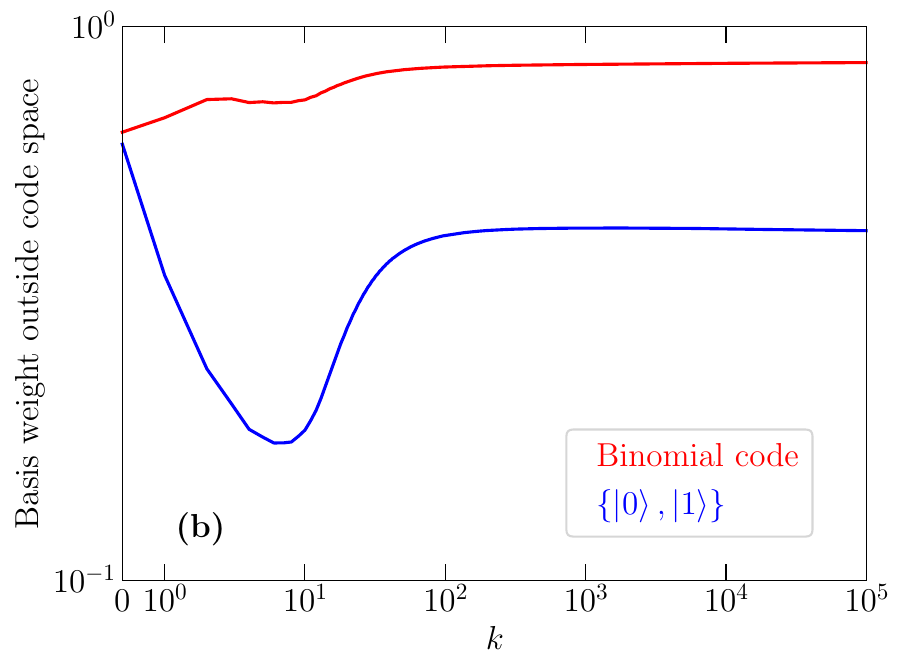}
\includegraphics[width=8.0cm]{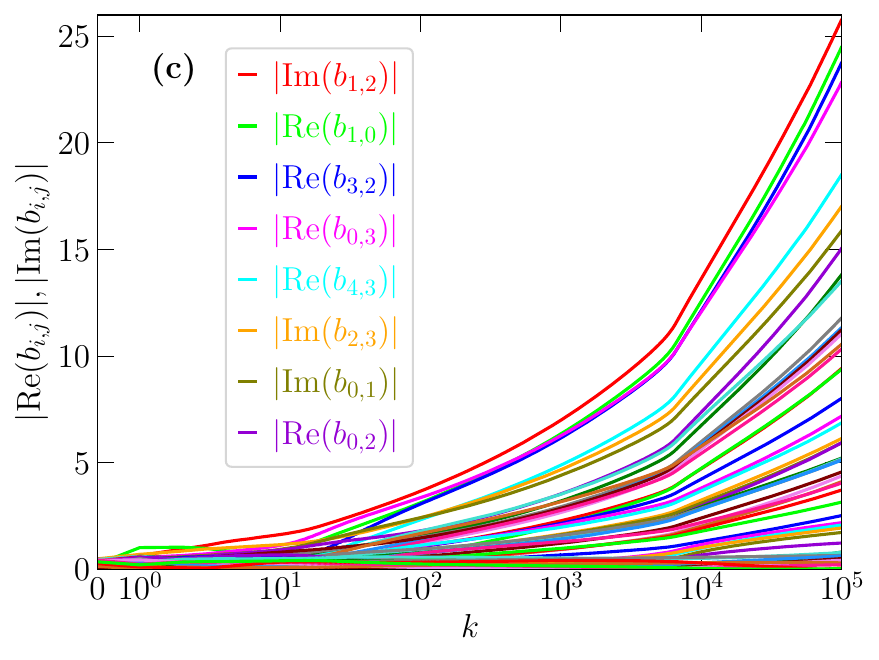}
\includegraphics[width=8.0cm]{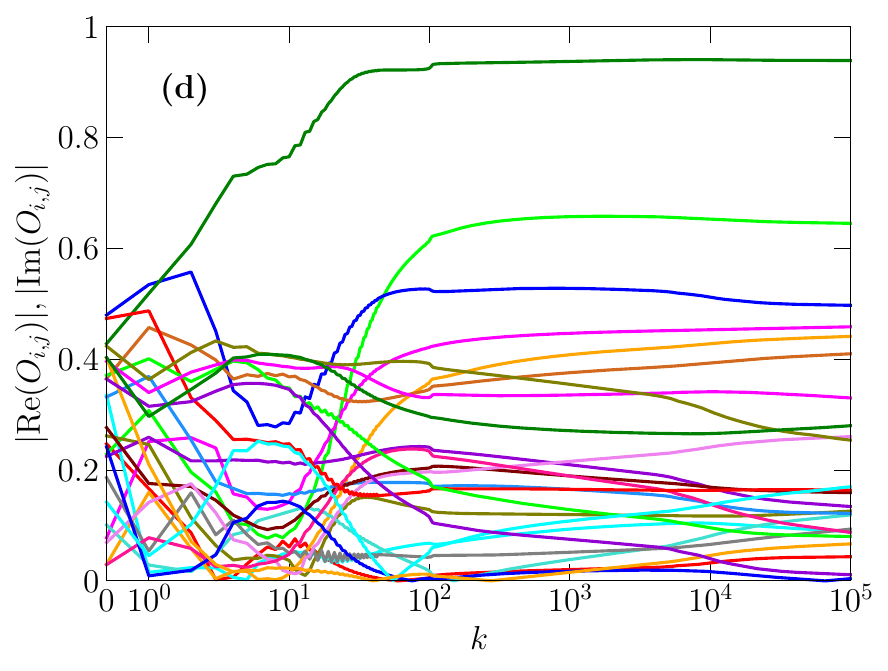}
\caption{Same as Fig.~\ref{Fig:Convergence501}, but for an instance in which the algorithm succeeded in finding an AQEC code with $F>F_0$. Such successful instances occurred in a few percent of the runs, each starting from a random seed. Although the fidelity is quite high ($F=0.998$) after $10^5$ iterations, Panel (b) clearly shows that the code is significantly different from the binomial code, which we expect is the optimal code in this model. Panel (a) suggests that there is still plenty of room to further optimize the code. On the other hand, the high fidelity indicates that this code is already a very good AQEC code. In some sense, this instance represents a case in which the algorithm discovered a new code that is very good, although it is still not as good as the binomial code, which has $1-F=6\times 10^{-6}$. As can be seen in Panel (c), this code requires that a large number of transitions are induced with varying transition rates. The legend specifies only the highest eight lines.}
\label{Fig:Convergence5}
\end{figure}

Next we consider the lowest five energy levels of a harmonic oscillator with photon loss decay, i.e.~the decay jump operator being given by the truncated annihilation operator
\begin{equation}
\hat{a} = \sqrt{\gamma} \left(
\begin{array}{ccccc}
0 & 1 & 0 & 0 & 0 \\
0 & 0 & \sqrt{2} & 0 & 0 \\
0 & 0 & 0 & \sqrt{3} & 0 \\
0 & 0 & 0 & 0 & 2 \\
0 & 0 & 0 & 0 & 0
\end{array}
\right).
\end{equation}
This system allows the implementation of the bosonic binomial code \cite{Michael}. The code space is spanned by the states $(\ket{0}+\ket{4})/\sqrt{2}$ and $\ket{2}$. The annihilation operator takes these states into the states $\ket{3}$ and $\ket{1}$, respectively, both with rate $2\gamma$.

One might intuitively think that the error correction is achieved by using the induced decay jump operator
\begin{equation}
\hat{b} = \sqrt{\Gamma} \left(
\begin{array}{ccccc}
0 & 0 & 0 & \frac{1}{\sqrt{2}} & 0 \\
0 & 0 & 0 & 0 & 0 \\
0 & 1 & 0 & 0 & 0 \\
0 & 0 & 0 & 0 & 0 \\
0 & 0 & 0 & \frac{1}{\sqrt{2}} & 0
\end{array}
\right).
\label{Eq:BinomialCatInducedDecayOp}
\end{equation}
This operator does indeed return the state after an error to the original state. However, one more error-correction component is needed. When no photon loss occurs during a time $\delta t$, the state $(\ket{0}+\ket{4})/\sqrt{2}$ is deformed and becomes $c(\ket{0}+e^{-\gamma\delta t}\ket{4})$, where $c$ is a normalization constant. To return this state to an equal superposition of $\ket{0}$ and $\ket{4})$, we must apply a correction operation. This correction can be achieved via the unitary operator
\begin{equation}
\hat{U} = \exp \left\{ -i \delta\theta \hat{\sigma}_y^{(04)} \right\},
\end{equation}
where $\delta\theta = \pi/2 - 2 \arctan\left( e^{-\gamma\delta t} \right)=\gamma\delta t + O(\delta t^2)$ and $\hat{\sigma}_y^{(04)} = i \left( \ket{4}\bra{0} - \ket{0}\bra{4} \right)$. This operation can in turn be implemented by adding to the Hamiltonian the time-independent operator $\hat{O}=\gamma\hat{\sigma}_y^{(04)}$. In other words, $O_{0,4}=-O_{4,0}=-i$, with all other matrix elements in $\hat{O}$ equal to zero. It should be noted that the operator $\hat{\sigma}_y^{(04)}$ involves a four-photon process and is therefore challenging to realize experimentally. There are, however, theoretical proposals and initial experimental results on implementing such multi-photon operations, especially in superconducting circuits \cite{Kwon2022,Chang,Eriksson}.

If we take the code space $\left\{ (\ket{0}+\ket{4})/\sqrt{2}, \ket{2} \right\}$ and the matrices $\hat{O}=\gamma\hat{\sigma}_y^{(04)}$ and $\hat{b}$ as in Eq.~(\ref{Eq:BinomialCatInducedDecayOp}) with $\Gamma/\gamma=10^6$, and we set $\gamma\tau=1$, we obtain the fidelity $F=0.999994=1-6 \gamma/\Gamma$. If we run the optimization algorithm using the above settings, the fidelity increases by less than $\sim 10^{-9}$ in $10^3$ iterations, at which point the algorithm meets our early termination condition.

If we fix $\hat{O}=\gamma\hat{\sigma}_y^{(04)}$ and $\hat{b}$ as in Eq.~(\ref{Eq:BinomialCatInducedDecayOp}) with $\Gamma/\gamma=10^6$, and we start the search from a randomly generated basis, the fidelity comes within $10^{-8}$ of the maximum value after about 10-15 iterations. In other words, the code basis converges quickly. Similarly, if we fix the code space to $\left\{ (\ket{0}+\ket{4})/\sqrt{2}, \ket{2} \right\}$ and $\hat{b}$ as in Eq.~(\ref{Eq:BinomialCatInducedDecayOp}) with $\Gamma/\gamma=10^6$, and we optimize $\hat{O}$, the fidelity quickly rises and saturates within about 10-15 iterations. The maximum fidelity is typically $10^{-6}$ lower than the maximum value of $F=0.999994$. As with the code basis, the Hamiltonian operator $\hat{O}$ converges quickly. If we fix the code space to $\left\{ (\ket{0}+\ket{4})/\sqrt{2}, \ket{2} \right\}$ and $\hat{O}=\gamma\hat{\sigma}_y^{(04)}$, and we optimize $\hat{b}$, the fidelity reaches $F=0.9998$ after $10^5$ iterations. When compared with the other two components of the AQEC code, the induced decay operator $\hat{b}$ is the slowest-converging component. In other words, the convergence behaviour of the individual AQEC code components is similar to that observed in the case of the ququart.

When we do not fix any of the code components and optimize all three of them with our standard initial guess conditions, we find that in the majority of the runs the fidelity converges within a few tens of iterations to the free-evolution value $F_0$ and terminates after a few thousand iterations. One example of such a run is illustrated in Fig.~\ref{Fig:Convergence501}. The fact that the fidelity converges to $F_0$ indicates that the algorithm is converging to the trivial solution with the code space $\left\{ \ket{0}, \ket{1} \right\}$ and $\hat{b}=\hat{O}=0$. This conclusion can also be made from inspecting the basis progression illustrated in Fig.~\ref{Fig:Convergence501}(b).

In a few percent of the runs, the fidelity exceeds $F_0$ within the first 100 iterations and goes on to reach $F\approx 0.998$ after $10^5$ iterations. One example of this situation is shown in Fig.~\ref{Fig:Convergence5}. The fact that different runs converged to different AQEC codes, with drastically different values of the fidelity, suggests that there are local optima in the fidelity landscape.

As can be seen in Fig.~\ref{Fig:Convergence5}(b-d), the code reached after $10^5$ iterations is still far from the binomial code. The slope of the fidelity in Fig.~\ref{Fig:Convergence5}(a) suggests that many more optimization iterations would be required to reach the level $1-F\sim 10^{-6}$. Furthermore, convergence towards the optimal code can be slowed down or even thwarted by the fact that the (mostly benign) admixture of error states in the dynamics translates into an increased penalty for using high photon number states. Nevertheless, this situation is a result of the fact that there are many vastly different AQEC codes with extremely high fidelities.

\section{Five-level system with perturbed-photon-loss decay law}

\begin{figure}[h]
\includegraphics[width=10.0cm]{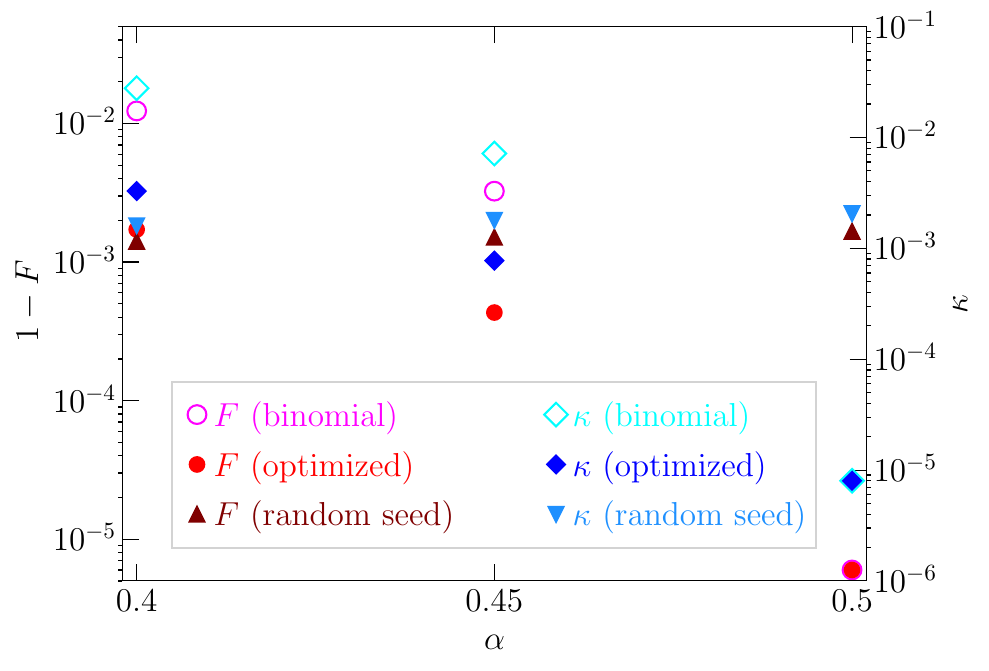}
\caption{Optimizing AQEC code for a slightly perturbed system. The infidelity $(1-F)$ and the decay rate suppression factor $\kappa$ are plotted as functions of the exponent $\alpha$ in Eq.~(\ref{Eq:FiveLevelGeneralizedAnnihilationOp}). The data labeled ``binomial'' are obtained by applying the binomial code in its original form to the perturbed system. The data labeled ``optimized'' are obtained by optimizing the AQEC code starting from the binomial code as an initial guess. For the data labeled ``random seed'', we took the best result out of ten runs using random seeds for each value of $\alpha$. When $\alpha=0.5$ (unperturbed case), the binomial code performs better than any AQEC code that we obtained numerically starting from a random seed. Furthermore, numerical optimization did not improve on the binomial code. When $\alpha=0.45$, applying the binomial code in its original form gives a relatively low fidelity. Starting from the binomial code and numerically optimizing it gives a better fidelity than any code obtained from a random seed. When $\alpha=0.4$ the best AQEC obtained from a random seed performs better than the code obtained by optimizing the binomial code, indicating that the perturbation is so large that the binomial code is no longer a useful starting point for optimization. These results demonstrate how our algorithm can find optimal AQEC codes that cannot be obtained by simple intuition about the system.}
\label{Fig:PowerLawPerturbationFidelity}
\end{figure}

We now consider a situation similar to that treated in Sec.~\ref{Sec:Results}.D, but with system parameters that are slightly different from those of an ideal harmonic oscillator. In a realistic setup, we do not expect the parameters to follow perfectly regular patterns. For example, in a superconducting circuit, there will inevitably by nonlinear terms that shift the Hamiltonian away from the perfect harmonic oscillator form. In such situations, intuition can at best produce an approximate error correction protocol. Numerical optimization is then particularly useful to optimize the protocol and achieve the best possible performance. It should be noted here that the perturbation can typically be characterized experimentally, such that we can know precisely what perturbation we are dealing with. In other words, we consider only static perturbations, not dynamic noise that is unpredictable in every run of the experiment.

For definiteness, we take a 5-level system and consider a decay jump operator that deviates slightly from a truncated harmonic oscillator annihilation operator. Specifically, we take
\begin{equation}
\hat{a} = \sqrt{\gamma} \left(
\begin{array}{ccccc}
0 & 1 & 0 & 0 & 0 \\
0 & 0 & 2^{\alpha} & 0 & 0 \\
0 & 0 & 0 & 3^{\alpha} & 0 \\
0 & 0 & 0 & 0 & 4^{\alpha} \\
0 & 0 & 0 & 0 & 0
\end{array}
\right),
\label{Eq:FiveLevelGeneralizedAnnihilationOp}
\end{equation}
with $\alpha$ slightly below 0.5. The results are shown in Fig.~\ref{Fig:PowerLawPerturbationFidelity}. If we take the binomial code in its original form and apply it to the cases $\alpha=0.45$ and $\alpha=0.4$, we obtain the fidelity values $F=0.9967$ and $F=0.988$, respectively. While these are high numbers, indicating that the binomial code is still a good AQEC code in the presence of the perturbation, the fidelity can be improved further with numerical optimization. We therefore take the original binomial code and use it as the initial guess for the automated code optimization algorithm. We now obtain the fidelity values $F=0.99957$ and $0.9983$, respectively. In both cases, the optimization leads to a reduction in the effective decay rate by a factor of about 9. In the case $\alpha=0.45$, starting the optimization from the binomial code produced the highest fidelity and lowest effective decoherence rate among all our calculations. In the case $\alpha=0.4$, the best code that we found was obtained from a random-seed search, indicating that the perturbation is so large that the binomial code does not provide a computational advantage as a starting point for the search.

\section{Discussion}

An important question when using numerical methods to optimize the operation of quantum systems is how the algorithm scales for large systems. Our algorithm performs a simulation of the time evolution of the open quantum system. This step will follow the scaling laws of simulating a quantum system on a classical computer. In general, it will suffer form the exponential scaling of required resources if we try to apply the algorithm to a many-qudit system. As a result, we can say that our algorithm is most suitable for single- or few qudit systems. This limitation is not a major obstacle to the applicability of the algorithm. As in the case of qubit-based error correction, if the errors on different qudits are uncorrelated, the multi-qubit error rate can be negligibly small, such that it can be ignored when developing the error correction strategy.

It is also worth mentioning that there are methods to extend the range of applicability of classical algorithms to quantum control problems. In particular, machine learning techniques can help speed up the AQEC protocol discovery process \cite{Foesel}. Another idea that is being discussed in similar contexts is the possible use of hybrid classical-quantum algorithms, where the quantum evolution is determined using a quantum processor while the optimization task is handled by a classical processor \cite{Bausch}. Our algorithm is well-suited for this approach. Yet another possible way to improve scaling is the use of probabilistic search methods. We recently showed that a probabilistic random search can dramatically enhance the ability of optimal-control methods to find quantum gate decompositions of few-qubit operations \cite{Ashhab2024}. Similar ideas could be tried in the AQEC code search problem.

In this work, we treated the induced decay matrix $\hat{b}$ as a control parameter to effect the desired error correction dynamics. In practice, controlled operations are typically applied using drive control fields. It is generally possible to engineer desired decay dynamics using a combination of ancillary quantum systems, their natural decay mechanisms and appropriately chosen drive control pulses. This design procedure is part of the field of quantum reservoir engineering, which is advancing rapidly in recent years \cite{Poyatos,Myatt,Murch}.

Furthermore, in this work, we assumed that all matrix elements are accessible and adjustable. In one case (Fig.~\ref{Fig:Convergence5}), we found an AQEC code that requires tuning a large number of transition rates in a five-level system. In practice, some transitions are more difficult to control than others. One simple example is the fact that the two-step transition $\ket{0}\to\ket{2}$ in a superconducting qutrit is typically much more difficult to drive than the single-step transitions $\ket{0}\to\ket{1}$ and $\ket{1}\to\ket{2}$ \cite{Yurtalan}. Some matrices $\hat{b}$ are therefore easier to engineer than others. One can assess the difficulty of implementing a certain AQEC code partly by evaluating the difficulty of engineering the required matrix elements. It is also possible to include constraints in the optimization procedure, such that difficult transition matrix elements are set to zero and therefore excluded from the optimization procedure. A related treatment was used in Ref.~\cite{Wang}.

We also note that we have made the approximation of Markovian dynamics. There are cases in which the decoherence dynamics is non-Markovian and a different description is necessary. While the mathematical description of non-Markovian dynamics is more complicated, one can search for error-correcting codes in this case as well by simulating the dynamics and trying to maximize the fidelity of the dynamical state with the initial (unknown) quantum state.

\section{Conclusion}
\label{Sec:Conclusion}

In conclusion, we have developed a method to identify and/or optimize AQEC codes in a general open quantum system. Once the system is characterized, the system parameters can be fed into the algorithm to obtain driving protocols that extend the lifetime of the quantum information. We have demonstrated the successful application of the algorithm to few-level systems. Our tests have shown cases in which the algorithm helps us obtain optimal, or near optimal, AQEC codes. At the same time, our analysis of the algorithm's performance has allowed us to identify possible weaknesses, convergence bottlenecks and areas where the algorithm could be improved. Our results on the performance of the algorithm can also serve as reference points for future algorithms designed to optimize AQEC codes. We expect that automated methods for the discovery and optimization of quantum error correction protocols like the one presented in this work will be valuable tools for achieving the best possible performance from quantum computing devices in the future.

\section*{Acknowledgment}

We would like to thank Uwe Fischer, Kouichi Semba and Fumiki Yoshihara for useful discussions. This work was supported by Japan's Ministry of Education, Culture, Sports, Science and Technology (MEXT) Quantum Leap Flagship Program Grant Number JPMXS0120319794.

\section*{Data availability}

The datasets generated and/or analysed during the current study are available from the author on reasonable request.

\appendix

\section{Optimizing individual components of AQEC code}
\label{Sec:AppendixOptimization}

\begin{figure}[h]
\includegraphics[width=8.0cm]{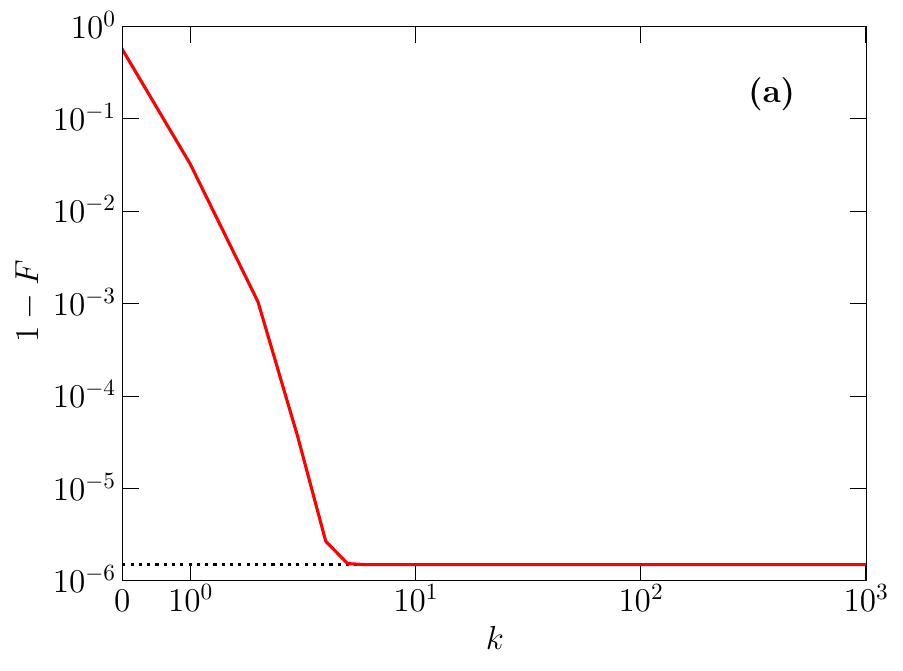}
\includegraphics[width=8.0cm]{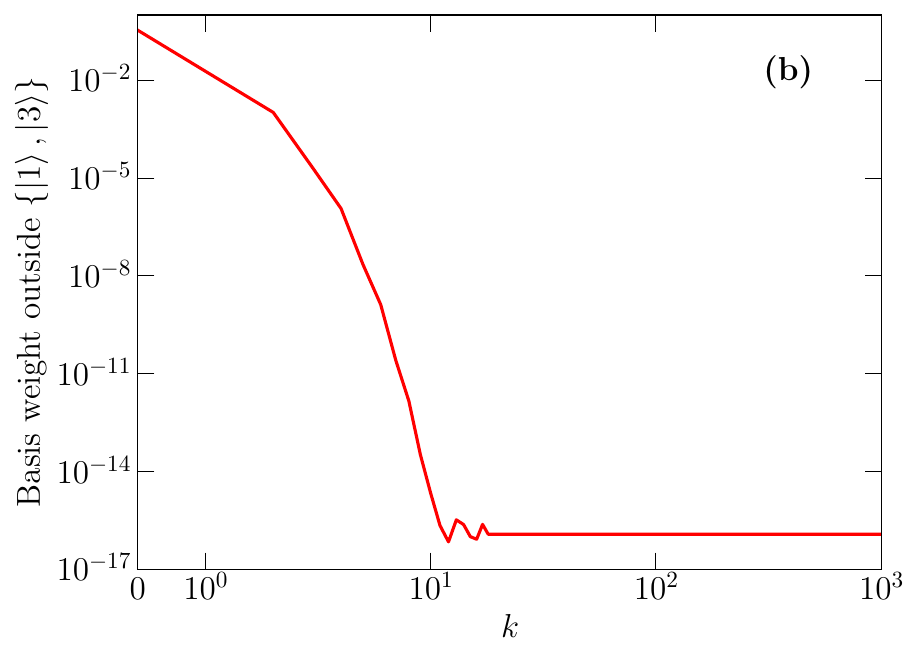}
\caption{Infidelity ($1-F$) and code space dependence on optimization iteration number $k$ for 4-level system with uniform decay rates. In this figure, we fix the matrices $\hat{b}$ and $\hat{O}$ at their optimal values for the 13 code described in the main text. The algorithm reaches an extremely good code basis within the first 10 iterations. The dotted line in Panel (a) represents the fidelity value $1-F=1.5\times 10^{-6}$, which is the expected value for the optimal code. These results indicate that optimizing the code space requires little computational time.}
\label{Fig:ConvergenceOptimizeBasis}
\end{figure}

\begin{figure}[h]
\includegraphics[width=8.0cm]{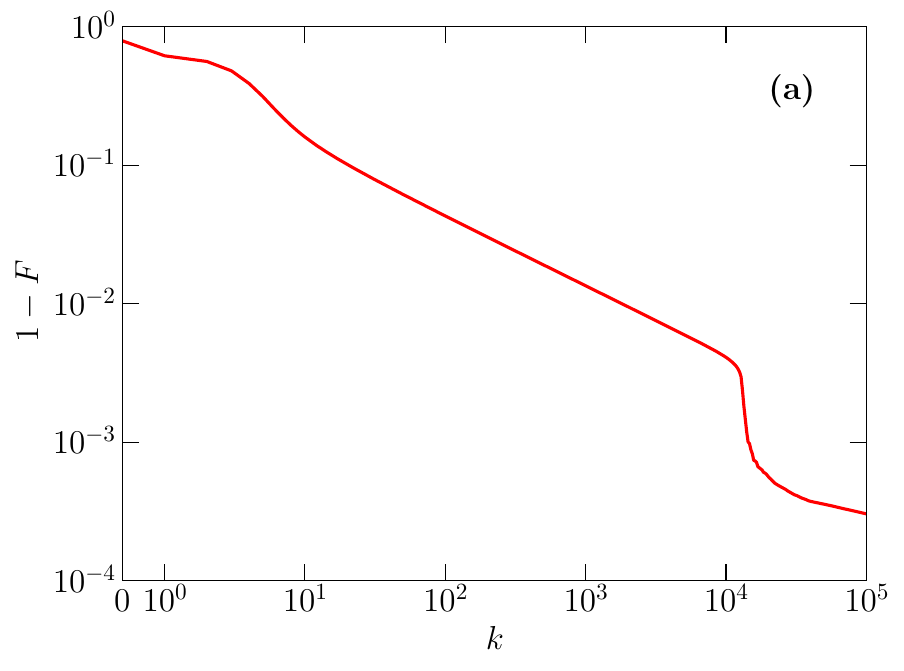}
\includegraphics[width=8.0cm]{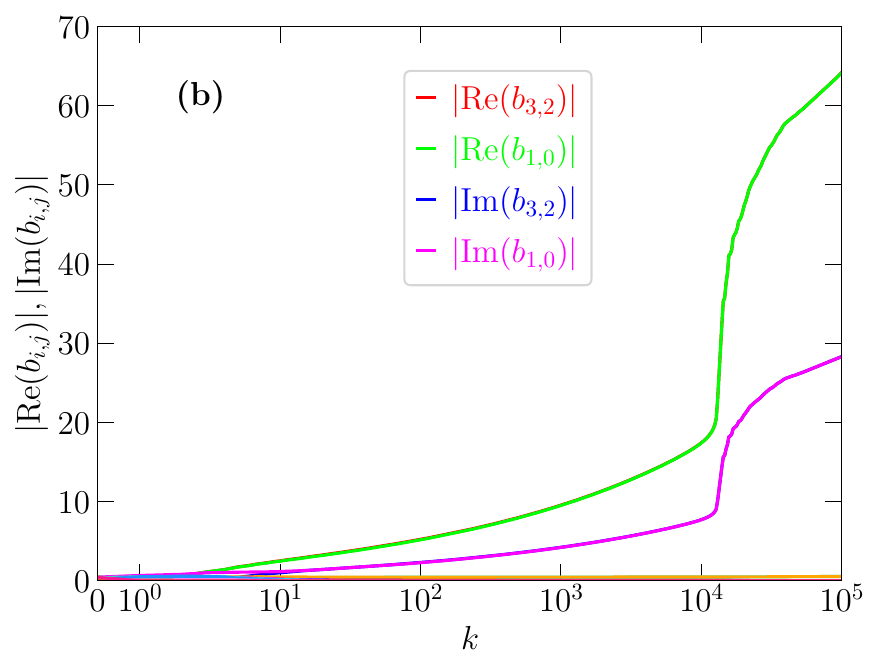}
\caption{Infidelity ($1-F$) and absolute values of the real and imaginary parts of the induced decay matrix elements $b_{i,j}$ as functions of optimization iteration number $k$ for 4-level system with uniform decay rates. In this figure, we fix the code space and matrix $\hat{O}$ at their optimal values in the 13 code. The fidelity reaches $F\approx 0.9997$ after $10^5$ iterations. The finite slope at $k=10^5$ in Panel (a) indicates that further improvement in the fidelity can be achieved with further optimization. In Panel (b), the legend specifies only the four lines that grow significantly as the optimization algorithm progresses. Each one of the two visible lines that reach high values is itself two lines that are extremely close to each other and are indistinguishable at the scale of this plot. These four lines correspond to the real and imaginary parts of $b_{1,0}$ and $b_{3,2}$. Depending on the random initial guess used in the search, in some computational runs the real part is higher, while in other runs the imaginary part is higher. The two matrix elements $b_{1,0}$ and $b_{3,2}$ are indeed expected to be equal with a large absolute value in the optimal code. These results show that the optimization algorithm is able to find the optimal matrix $\hat{b}$, but the convergence is much slower than that of the code space optimization.}
\label{Fig:ConvergenceOptimizeDecay}
\end{figure}

\begin{figure}[h]
\includegraphics[width=8.0cm]{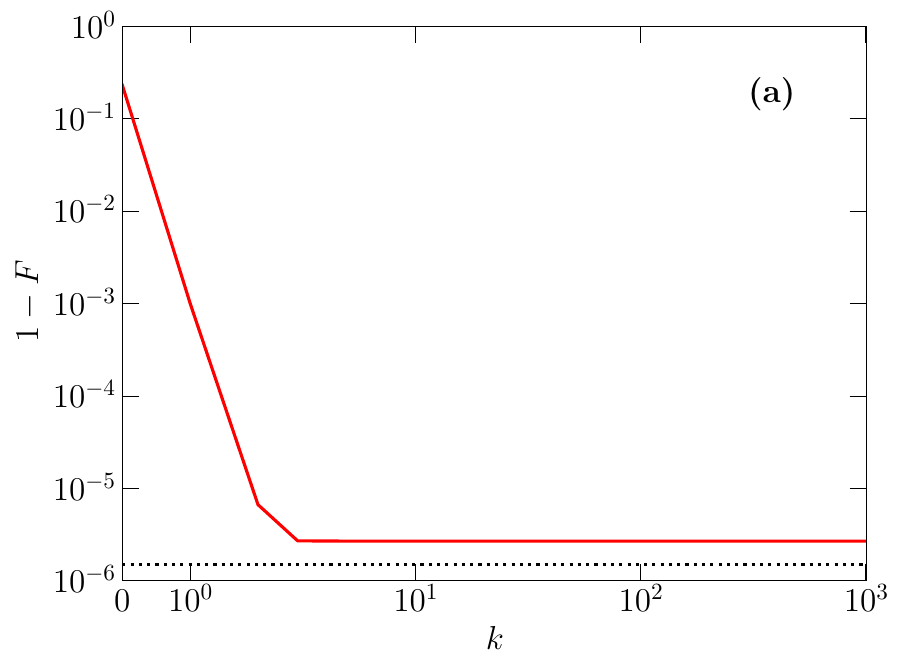}
\includegraphics[width=8.0cm]{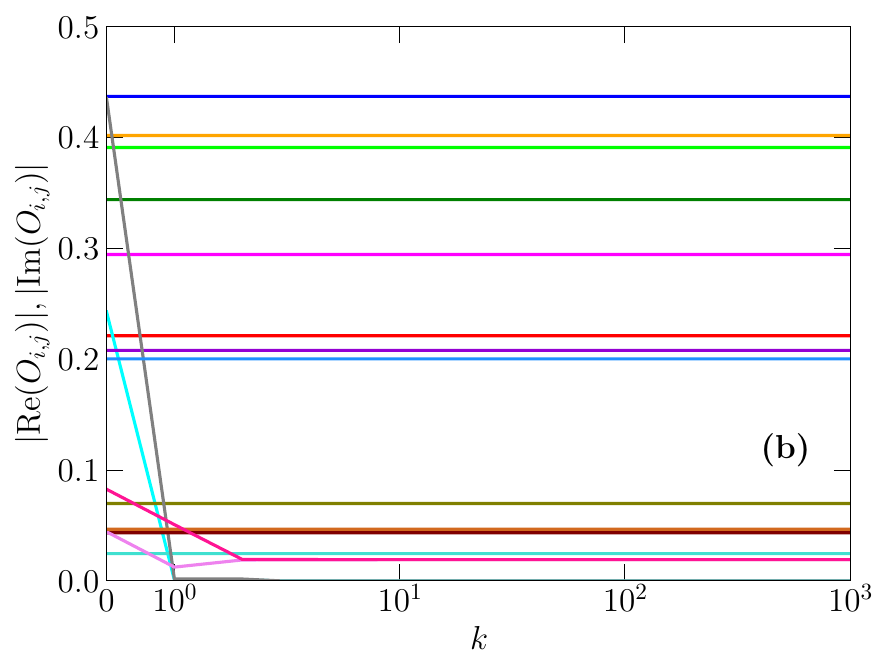}
\caption{Infidelity ($1-F$) and absolute values of the real and imaginary parts of the coherent coupling matrix elements $O_{i,j}$ as functions of optimization iteration number $k$ for 4-level system with uniform decay rates. In this figure, we fix the code space at the optimal choice $\left\{ \ket{1}, \ket{3} \right\}$ and set $\hat{b}$ as in Eq.~(\ref{Eq:FourLevelInducedDecayOp}) with $\Gamma/\gamma=10^6$. The dotted line in Panel (a) represents the infidelity value $1-F=1.5\times 10^{-6}$, which is the expected value for the optimal choice of $\hat{O}$. The infidelity goes below $3\times 10^{-6}$ after just a few iterations and barely changes after that. The real and imaginary parts of the matrix element $O_{1,3}$ (and hence also $O_{3,1}$) shrink almost to zero after just one iteration, the matrix element $O_{1,1}$ and $O_{3,3}$ become equal within a few iterations, while all other matrix elements barely change in $10^3$ iterations, indicating that most matrix elements have almost no impact on the performance of the AQEC code. As with the code space, convergence is very fast.}
\label{Fig:ConvergenceOptimizeCoupling}
\end{figure}

In this Appendix, we present additional results that elucidate the performance of the optimization algorithm in optimizing the different components of the AQEC code. We use the same 4-level model as in Sec.~\ref{Sec:Results}.B. We perform a few different optimization tasks, focusing on one component at a time.

First, we fix $\hat{O}=0$ and $\hat{b}$ as in Eq.~(\ref{Eq:FourLevelInducedDecayOp}) with $\Gamma/\gamma=10^6$, and we search for the optimal code space, starting the search from a randomly generated basis. The search converges to the code space $\left\{ \ket{1}, \ket{3} \right\}$, and the fidelity comes within $10^{-8}$ of the maximum value, after only a few optimization iterations, as shown in Fig.~\ref{Fig:ConvergenceOptimizeBasis}.

Next we fix the code words at the states $\left\{ \ket{1}, \ket{3} \right\}$ and $\hat{O}=0$, and we run the algorithm to optimize $\hat{b}$. The results of a typical run are shown in Fig.~\ref{Fig:ConvergenceOptimizeDecay}. As explained in Sec.~\ref{Sec:Algorithm}, $\hat{b}$ is initially filled with complex random numbers whose real and imaginary parts are each chosen from a uniform distribution in the range $[-0.5,0.5]$. The search typically reaches a fidelity of $F\approx 0.996$ after $10^4$ iterations and $F\approx 0.99969$ after $10^5$ iterations. The matrix $\hat{b}$ converges to a matrix where the matrix elements $b_{1,0}$ and $b_{3,2}$ are almost equal and have a large absolute value. After $10^5$ iterations, the absolute value of $b_{1,0}$ is typically on the order of $10^2$, the relative difference between $b_{1,0}$ and $b_{3,2}$ is on the order of $10^{-4}$, and the next largest matrix element is two orders of magnitude smaller than $|b_{1,0}|$. The slow convergence in this case can, at least partly, be explained by the fact that the optimal code described in Sec.~\ref{Sec:Results}.B requires having $b_{1,0}=b_{3,2}$ and $\left| b_{1,0}\right|\gg\gamma$, with the other matrix elements being equal to zero. The code does indeed evolve in this direction as the optimization algorithm progresses. However, the inequality between $b_{1,0}$ and $b_{3,2}$ during the optimization creates a bottleneck for convergence. By considering the fidelity landscape, we can see that a small difference between $b_{1,0}$ and $b_{3,2}$ reduces the fidelity much more than any fidelity increase that can be gained by increasing the two values together by $|b_{1,0}-b_{3,2}|$. As a result, the algorithm identifies the optimal update policy as bringing the two values closer to each other, while slightly increasing the average. While the matrix elements do move in the correct direction when observed at the scale of many iterations, the approach to the optimal situation with the correct relations between the different matrix elements is rather inefficient and causes slow convergence.

Next, we fix the code words at the states $\left\{ \ket{1}, \ket{3} \right\}$ and $\hat{b}$ as in Eq.~(\ref{Eq:FourLevelInducedDecayOp}) with $\Gamma/\gamma=10^6$. We optimize the operator $\hat{O}$, which we expect to have the optimal value $\hat{O}=0$. The results of a typical run are shown in Fig.~\ref{Fig:ConvergenceOptimizeCoupling}. The fidelity consistently reaches a value between $0.999997$ and $0.999998$ after just one optimization iterations and remains constant (up to 12 significant figures) after only a few iterations. As a result, the search terminates as soon as the other early-termination conditions are satisfied, i.e.~after about $10^3$ iterations. In most runs, the matrix elements $O_{1,3}$ and $O_{3,1}$ shrink to at most $\sim 1 \times 10^{-6}$, the matrix elements $O_{1,1}$ and $O_{3,3}$ become equal, while all other matrix elements barely change from their initial random values, with net changes consistently below $10^{-4}$. The reason why only four matrix elements converge towards their optimal values is that all the other matrix elements involve states that remain almost unoccupied and therefore have a negligible effect on the fidelity. Only the matrix elements $O_{1,1}$, $O_{3,3}$, $O_{1,3}$ and $O_{3,1}$ cause a significant change in a general state in the code space. Even the matrix elements that mix states in the code space with states outside the code space (e.g.~$O_{0,1}$) are effectively suppressed by the large induced decay described by the matrix $\hat{b}$. As is common in optimization problems, when a certain parameter has a negligible effect on the cost function (i.e.~the fidelity in our case), the algorithm does not prioritize optimizing the parameter. For this reason, most matrix elements in $\hat{O}$ are barely updated and remain very close to their initial values. It is somewhat surprising that this stagnation in suppressing the almost irrelevant matrix elements occurs even when the fidelity is still $\sim 10^{-6}$ away from the maximum possible value, which is not an extremely small difference.

In about 10\% of the calculations, the matrix elements $O_{1,1}$, $O_{3,3}$, $O_{1,3}$ and $O_{3,1}$ did not converge towards the optimal values described in the previous paragraph. For example, if we fix the matrix elements $O_{1,1}=O_{3,3}=0$, the matrix elements $O_{1,3}$ and $O_{3,1}$ occasionally did not converge towards zero, but towards values that are integer multiples of $2\pi\gamma$ in magnitude. The reason is as follows: the dynamics induced by these matrix elements is coherent and does not represent irreparable decoherence in the quantum state, provided that the matrix elements are known. Therefore, in principle, these matrix elements can take any values without contributing to decoherence. However, our optimization algorithm requires that the state remain at its initial value. Therefore, $O_{1,1}$, $O_{3,3}$, $O_{1,3}$ and $O_{3,1}$ must converge to a combination of values that takes the code words back to their initial values when $\gamma\tau=1$, which is the time setting at which we evaluate the fidelity.

\section{General qudit with uniform decay rates}
\label{Sec:AppendixGeneralQudit}

It is not difficult to find an intuitive generalization of the 13 code for an $n$-level system with $n>4$. In particular, if $n$ is even, the code space can be taken as $\left\{ \ket{n/2}, \ket{n} \right\}$. The loss of $p$ excitations takes the state to the space $\left\{ \ket{n/2-p}, \ket{n-p} \right\}$, without degrading the quantum coherence in the state. These errors can be corrected by the application of multiple induced-decay jump operators (one for each value of $p$), or alternatively the application of a single jump operator that corrects all errors. Each one of the separate induced-decay jump operators would fix an error associated with the loss of $p$ excitations, such that the quantum state returns to the code space by adding $p$ excitations, without affecting the quantum coherence between the states $\ket{n/2-p}$ and $\ket{n-p}$. It is worth noting here that one intuitive candidate for a single jump operator that would corrects all errors, namely the one obtained by setting $b_{n/2,n/2-p}=b_{n,n-p}$ for $1 \leq p \leq n/2$ with zero matrix elements otherwise, does not perform the desired task, because it merges states that had previously incurred different numbers of errors. As a result, it does not provide a mechanism to get rid of the entropy that is created in the system when the different errors occur. If instead we set $b_{n/2-p+1,n/2-p}=b_{n-p+1,n-p}$ for $1 \leq p \leq n/2$, we obtain a good AQEC code. For example, if we take $n=6$ and $\Gamma/\gamma=10^6$ for both $p=1$ and $p=2$, and we set $\gamma\tau =1$, we numerically obtain a fidelity of $F=0.999999=1-1.0\times 10^{-6}$. The small reduction away from $F=1$ can now almost completely be accounted for by the admixture of error states in the long-time dynamics, with no noticeable contribution from the long-time decoherence of the code words.

We ran the optimization algorithm on the case of a 6-level system starting from a random seed, i.e.~random guess for the AQEC code. In the majority of the runs, the fidelity reached $F\approx 0.9992$ after $10^5$ iterations. The maximum value reached in one run (out of ten runs) was $F=0.99938$. These fidelity values are much higher than the free-evolution value $F_0$ given in Sec.~\ref{Sec:Results}.A, but short of the value 0.999999 obtained with intuitive reasoning above. The optimization algorithm is clearly able to find good AQEC protocols. We expect that it will converge to the optimal code if it is run longer. One caveat here is that there are a large number of extremely good AQEC codes in this 6-level system, such that the algorithm can already reach fidelity values at the level of several nines with codes that are drastically different from the possibly optimal one described above. Furthermore, convergence tends to be slow when the fidelity approaches the maximum, mainly because of the slow convergence of the induced decay matrix $\hat{b}$. An additional complication that does not affect this particular example but can be important for other decay models is the fact that higher levels generally decay faster and result in higher admixtures of error states. Even if these admixtures do not affect the effective decay rate, they contribute a small reduction to the fidelity, making it difficult for the algorithm to distinguish between many different codes that all have extremely high fidelities.

\section{Ququart with general power decay law}
\label{Sec:AppendixVariableAlpha}

\begin{figure}[h]
\includegraphics[width=10.0cm]{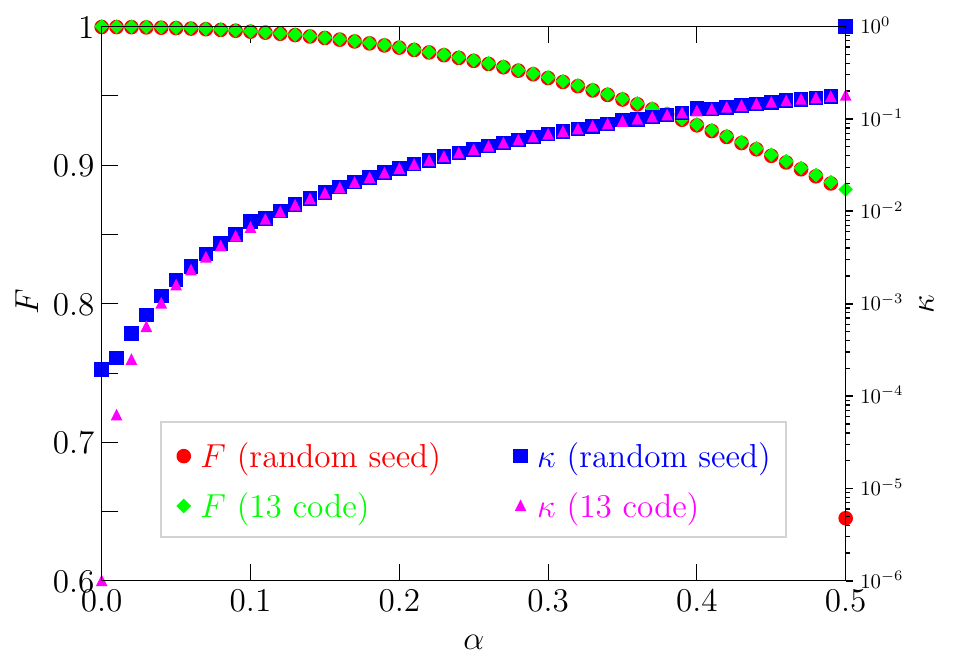}
\caption{Code fidelity $F$ and decay rate suppression factor $\kappa$ for the optimal AQEC code for a 4-level system with decay jump operator given by Eq.~(\ref{Eq:FourLevelGeneralizedAnnihilationOp}). In one set of calculations, we start the search from the 13 code, and in another set we use a random seed. The case $\alpha=0$ corresponds to uniform decay rates, where the 13 code is expected to be optimal. The case $\alpha=0.5$ corresponds to photon loss decay. Interestingly, starting the search from a random seed did not produce any nontrivial codes for $\alpha=0.5$, but nontrivial codes are found with high probability as soon as we move away from $\alpha=0.5$. Furthermore, using the 13 code as a starting point for the search produces a nontrivial code even at $\alpha=0.5$. These results represent an example in which our algorithm finds AQEC codes in cases where there are no obvious intuitive codes.}
\label{Fig:PowerLawDecayVariableAlpha}
\end{figure}

To further investigate the difference between the uniform decay case treated in Sec.~\ref{Sec:Results}.B and the photon loss case treated in Sec.~\ref{Sec:Results}.C, we ask what happens in intermediate cases. In particular, we consider the jump operator
\begin{equation}
\hat{a} = \sqrt{\gamma} \left(
\begin{array}{cccc}
0 & 1 & 0 & 0 \\
0 & 0 & 2^{\alpha} & 0 \\
0 & 0 & 0 & 3^{\alpha} \\
0 & 0 & 0 & 0
\end{array}
\right),
\label{Eq:FourLevelGeneralizedAnnihilationOp}
\end{equation}
where the exponent $\alpha$ is between 0 (uniform decay case) and 0.5 (photon loss case). The fidelity as a function of $\alpha$ is plotted in Fig.~\ref{Fig:PowerLawDecayVariableAlpha}. For the random-seed data, we ran the algorithm 5 times for each setting and chose the highest fidelity, noting here that most runs produced essentially the same results. Interestingly, even for $\alpha=0.49$, the algorithm consistently finds a nontrivial AQEC code with fidelity $F\approx 0.8867$, which is close to the value mentioned above when applying the 13 code. Indeed, if we inspect the codes found by the algorithm for all values of $\alpha$ excluding 0.5, we find that the code space is approximately $\left\{ \ket{1}, \ket{3} \right\}$. If we start the search from the 13 code, we find that the fidelity agrees with that found by the random-seed search for most values of $\alpha$. One exception is the case $\alpha=0.5$, where the random-seed search fails to find the optimal code. Another exception is the case of small $\alpha$, where starting the search from a very good guess (i.e.~the 13 code) gives a noticeably higher fidelity than the random-seed search for the same number of optimization iterations.


\begin{thebibliography}{99}

\bibitem{Ladd} T. D. Ladd, F. Jelezko, R. Laflamme, Y. Nakamura, C. Monroe, and J. L. O'Brien, Quantum computers, Nature {\bf 464}, 45 (2010).

\bibitem{Buluta} I. Buluta, S. Ashhab, and F. Nori, Natural and artificial atoms for quantum computation, Rep. Prog. Phys. {\bf 74}, 104401 (2011).

\bibitem{Shor} P. W. Shor, Scheme for reducing decoherence in quantum computer memory, Phys. Rev. A {\bf 52}, 2493 (1995).

\bibitem{Steane} A. M. Steane, Error correcting codes in quantum theory, Phys. Rev. Lett. {\bf 77}, 793 (1996).

\bibitem{Bennett} C. H. Bennett, D. P. DiVincenzo, J. A. Smolin, and W. K. Wootters, Mixed state entanglement and quantum error correction, Phys. Rev. A {\bf 54}, 3824 (1996).

\bibitem{Knill} E. Knill and R. Laflamme, Theory of quantum error-correcting codes, Phys. Rev. A {\bf 55}, 900 (1997).

\bibitem{Nielsen} M. A. Nielsen and I. L. Chuang, {\it Quantum Computation and Quantum Information} (Cambridge University Press, New York, 2000).

\bibitem{Terhal} B. M. Terhal, Quantum error correction for quantum memories, Rev. Mod. Phys. {\bf 87}, 307 (2015).

\bibitem{Girvin} S. M. Girvin, Introduction to quantum error correction and fault tolerance, SciPost Phys. Lect. Notes 70 (2023).

\bibitem{Barnes} J. P. Barnes and W. S. Warren, Automatic quantum error correction, Phys. Rev. Lett. {\bf 85}, 856 (2000).

\bibitem{Sarovar} M. Sarovar and G. J. Milburn, Continuous quantum error correction by cooling, Phys. Rev. A {\bf 72}, 012306 (2005).

\bibitem{Campbell} E. T. Campbell, Enhanced fault-tolerant quantum computing in d-level systems, Phys. Rev. Lett. {\bf 113}, 230501 (2014).

\bibitem{Lihm} J.-M. Lihm, K. Noh, and U. R. Fischer, Implementation-independent sufficient condition of the Knill-Laflamme type for the autonomous protection of logical qudits by strong engineered dissipation, Phys. Rev. A {\bf 98}, 012317 (2018).

\bibitem{Lebreuilly} J. Lebreuilly, K. Noh, C.-H. Wang, S. M. Girvin, and L. Jiang, Autonomous quantum error correction and quantum computation, arXiv:2103.05007.

\bibitem{Kwon2025} H. Kwon, U. R. Fischer, S.-W. Lee, L. Jiang, Restoring Heisenberg scaling in time via autonomous quantum error correction, arXiv:2504.13168.

\bibitem{Kapit} E. Kapit, Hardware-efficient and fully autonomous quantum error correction in superconducting circuits, Phys. Rev. Lett. {\bf 116}, 150501 (2016).

\bibitem{Li2024} Z. Li, T. Roy, D. Rodr\'iguez P\'erez, K.-H. Lee, E. Kapit, and D. I. Schuster, Autonomous error correction of a single logical qubit using two transmons, Nature Commun. {\bf 15}, 1681 (2024).

\bibitem{Kwon2022} S. Kwon, S. Watabe, and J.-S. Tsai, Autonomous quantum error correction in a four-photon Kerr parametric oscillator, npj Quantum Inf. {\bf 8}, 40 (2022).

\bibitem{Gertler} J. M. Gertler, B. Baker, J. Li, S. Shirol, J. Koch, and C. Wang, Protecting a bosonic qubit with autonomous quantum error correction, Nature {\bf 590}, 243 (2021).

\bibitem{LachanceQuirion} D. Lachance-Quirion, M.-A. Lemonde, J. O. Simoneau, L. St-Jean, P. Lemieux, S. Turcotte, W. Wright, A. Lacroix , J. Fr\'echette-Viens , R. Shillito, F. Hopfmueller, M. Tremblay, N. E. Frattini, J. Camirand Lemyre, and P. St-Jean, Autonomous quantum error correction of Gottesman-Kitaev-Preskill states, Phys. Rev. Lett. {\bf 132}, 150607 (2024).

\bibitem{Brock} B. L. Brock, S. Singh, A. Eickbusch, V. V. Sivak, A. Z. Ding, L. Frunzio, S. M. Girvin, and M. H. Devoret, Quantum error correction of qudits beyond break-even, Nature {\bf 641}, 612 (2025).

\bibitem{Leghtas} Z. Leghtas, S. Touzard, I. M. Pop, A. Kou, B. Vlastakis, A. Petrenko, K. M. Sliwa, A. Narla, S. Shankar, M. J. Hatridge, {\it et al.}, Confining the state of light to a quantum manifold by engineered two-photon loss, Science {\bf 347}, 853 (2015).

\bibitem{DeBry} K. DeBry, N. Meister, A. Valdes Martinez, C. D. Bruzewicz, X. Shi, D. Reens, R. McConnell, I. L. Chuang, and J. Chiaverini, Error correction of a logical qubit encoded in a single atomic ion, arXiv:2503.13908.

\bibitem{Li2025} Y. Li, Q. Mei, Q.-X. Jie, W. Cai, Y. Li, Z. Liu, Z.-J. Chen, Z. Xie, X. Cheng, X. Zhao, Z. Luo, M. Zhang, X.-B. Zou, C.-L. Zou, Y. Lin, and J. Du, Beating the break-even point with autonomous quantum error correction, arXiv:2504.16746.

\bibitem{Foesel} T. F\"osel, P. Tighineanu, T. Weiss, and F. Marquardt, Reinforcement learning with neural networks for quantum feedback, Phys. Rev. X {\bf 8}, 031084 (2018).

\bibitem{Wang} Z. Wang, T. Rajabzadeh, N. Lee, and A. H. Safavi-Naeini, Automated discovery of autonomous quantum error correction schemes, PRX Quantum 3, 020302 (2022).

\bibitem{Zeng} Y. Zeng, Z.-Y. Zhou, E. Rinaldi, C. Gneiting, and F. Nori, Approximate autonomous quantum error correction with reinforcement learning, Phys. Rev. Lett. {\bf 131}, 050601 (2023).

\bibitem{Reimpell} M. Reimpell and R. F. Werner, Iterative optimization of quantum error correcting codes, Phys. Rev. Lett. {\bf 94}, 080501 (2005).

\bibitem{Yamamoto2005} N. Yamamoto, S. Hara, and K. Tsumura, Suboptimal quantum-error-correcting procedure based on semidefinite programming, Phys. Rev. A {\bf 71}, 022322 (2005).

\bibitem{Fletcher2007} A. S. Fletcher, P. W. Shor, and M. Z. Win, Optimum quantum error recovery using semidefinite programming, Phys. Rev. A {\bf 75}, 012338 (2007).

\bibitem{Yamamoto2007} N. Yamamoto and M. Fazel, Computational approach to quantum encoder design for purity optimization, Phys. Rev. A {\bf 76}, 012327 (2007).

\bibitem{Kosut} R. L. Kosut, A. Shabani, and D. A. Lidar, Robust quantum error correction via convex optimization, Phys. Rev. Lett. {\bf 100}, 020502 (2008).

\bibitem{Fletcher2008} A. S. Fletcher, P. W. Shor, and M. Z. Win, Structured near-optimal channel-adapted quantum error correction, Phys. Rev. A {\bf 77}, 012320 (2008).

\bibitem{PoulsenNautrup} H. Poulsen Nautrup, N. Delfosse, V. Dunjko, H. J. Briegel, and N. Friis, Optimizing quantum error correction codes with reinforcement learning, Quantum {\bf 3}, 215 (2019).

\bibitem{Bausch} J. Bausch, A. W. Senior, F. J. H. Heras, T. Edlich, A. Davies, M. Newman, C. Jones, K. Satzinger, M. Y. Niu, S. Blackwell, G. Holland, D. Kafri, J. Atalaya, C. Gidney, D. Hassabis, S. Boixo, H. Neven, and P. Kohli, Learning high-accuracy error decoding for quantum processors, Nature {\bf 635}, 834 (2024).

\bibitem{Olle} J. Olle, R. Zen, M. Puviani, and F. Marquardt, Simultaneous discovery of quantum error correction codes and encoders with a noise-aware reinforcement learning agent, npj Quantum Inf. {\bf 10}, 126 (2024).

\bibitem{Su} V. P. Su, C. Cao, H.-Y. Hu, Y. Yanay, C. Tahan, and B. Swingle, Discovery of optimal quantum error correcting codes via reinforcement learning, Phys. Rev. Applied {\bf 23}, 034048 (2025).

\bibitem{Casanova} M. Casanova, K. Ohki, and F. Ticozzi, Finding quantum codes via Riemannian optimization, Quantum Sci. Technol. {\bf 10}, 025027 (2025).

\bibitem{Leung} D. W. Leung, M. A. Nielsen, I. L. Chuang, and Y. Yamamoto, Approximate quantum error correction can lead to better codes, Phys. Rev. A {\bf 56}, 2567 (1997).

\bibitem{Beny} C. B\'eny, and O. Oreshkov, General Conditions for Approximate Quantum Error Correction and Near-Optimal Recovery Channels, Phys. Rev. Lett. {\bf 104}, 120501 (2010).

\bibitem{Ng} H. K. Ng, and P. Mandayam, Simple approach to approximate quantum error correction based on the transpose channel, Phys. Rev. A {\bf 81}, 062342 (2010).

\bibitem{Ashhab2022} S. Ashhab, F. Yoshihara, T. Fuse, N. Yamamoto, A. Lupascu, and K. Semba, Speed limits for two-qubit gates with weakly anharmonic qubits, Phys.~Rev.~A {\bf 105}, 042614 (2022).

\bibitem{Michael} M. H. Michael, M. Silveri, R. T. Brierley, V. V. Albert, J. Salmilehto, L. Jiang, and S. M. Girvin, New class of quantum error-correcting codes for a bosonic mode, Physical Review X {\bf 6}, 031006 (2016).

\bibitem{Chang} C. W. S. Chang, C. Sabín, P. Forn-Díaz, F. Quijandría, A. M. Vadiraj, I. Nsanzineza, G. Johansson, and C. M. Wilson, Observation of three-photon spontaneous parametric down-conversion in a superconducting parametric cavity, Phys. Rev. X {\bf 10}, 011011 (2020).

\bibitem{Eriksson} A. M. Eriksson, T. Sépulcre, M. Kervinen, T. Hillmann, M. Kudra, S. Dupouy, Y. Lu, M. Khanahmadi, J. Yang, C. Castillo-Moreno, P. Delsing, and S. Gasparinetti, Universal control of a bosonic mode via drive-activated native cubic interactions, Nature Commun. {\bf 15}, 2512 (2024).

\bibitem{Ashhab2024} S. Ashhab, F. Yoshihara, M. Tsuji, M. Sato, and K. Semba, Quantum circuit synthesis via a random combinatorial search, Phys. Rev. A {\bf 109}, 052605 (2024).

\bibitem{Poyatos} J. F. Poyatos, J. I. Cirac, and P. Zoller, Quantum reservoir engineering with laser cooled trapped ions, Phys. Rev. Lett. {\bf 77}, 4728 (1996).

\bibitem{Myatt} C. J. Myatt, B. E. King, Q. A. Turchette, C. A. Sackett, D. Kielpinski, W. M. Itano, C. Monroe, and D. J. Wineland, Decoherence of quantum superpositions through coupling to engineered reservoirs, Nature (London) {\bf 403}, 269 (2000).

\bibitem{Murch} K. W. Murch, S. J. Weber, K. M. Beck, E. Ginossar, and I. Siddiqi, Reduction of the radiative decay of atomic coherence in squeezed vacuum, Nature (London) {\bf 499}, 62 (2013).

\bibitem{Yurtalan} M. A. Yurtalan, J. Shi, M. Kononenko, A. Lupascu, and S. Ashhab, Implementation of a Walsh-Hadamard gate in a superconducting qutrit, Phys. Rev. Lett. {\bf 125}, 180504 (2020).


\end{thebibliography}
\end{document}